\documentclass[twoside, a4paper,11pt]{article}
\usepackage[latin1]{inputenc}
\usepackage[T1]{fontenc}
\usepackage{graphicx}
\usepackage{times}
\usepackage{amssymb}

\newcommand{\be}{\begin{equation}}
\newcommand{\ee}{\end{equation}}

\title{Statistical analysis of Gene and Intergenic DNA Sequences}

\author{D.~Kugiumtzis \\
{\em Department of Mathematical, Physical and Computational Sciences} \\ 
{\em Polytechnic School, Aristotle University of Thessaloniki} \\
{\em Thessaloniki 54006, Greece} \\[5mm]
A.~Provata \\ 
{\em Institute of Physical Chemistry}\\ 
{\em National Research Center ``Demokritos''} \\ 
{\em Athens 15310, Greece}}

\begin{document}
\maketitle

\bibliographystyle{unsrt}

\begin{abstract}
Much of the on-going statistical analysis of DNA sequences is focused on the 
estimation of characteristics of coding and non-coding regions that would 
possibly allow discrimination of these regions. 
In the current approach, we concentrate specifically on genes and
intergenic regions. To estimate the level and type of correlation in these
regions we apply various statistical methods inspired from nonlinear time series
analysis, namely the probability distribution of tuplets, the Mutual Information
and the Identical Neighbour Fit. The methods
are suitably modified to work on symbolic sequences
and they are first tested for validity on sequences obtained
 from well--known simple deterministic and stochastic models.
Then they are applied to the DNA sequence of chromosome 1 of
{\em arabidopsis thaliana}. 
The results suggest that correlations do exist in the DNA sequence but 
they are weak and that intergenic sequences tend to be more correlated
than gene sequences. 
The use of statistical tests with surrogate data establish these 
findings in a rigorous statistical manner.  
\end{abstract}

\section{Introduction}

The goal of the ongoing genome projects is to detect and extract the
genetic information from the DNA sequences. 
In addition to any biological approaches, that constitute major 
contributions towards this goal, statistical analysis of DNA sequences 
has its own merit and has met tremendous interest in the recent years. 
In particular, much effort is focused on the estimation of 
statistical properties of coding and non-coding 
regions as well as on the discrimination of these regions. 
Starting from the pioneer work in \cite{Peng92} on the so-called ``DNA walk'' 
where long-range power law scaling was found on non-coding regions, a number
of correlation measures were used, such as power spectrum 
(e.g. see \cite{Buldyrev95}), detrended fluctuation analysis 
\cite{Peng94}, redundancy and entropy \cite{Ebeling91,Stanley99b}, and
mutual information \cite{Herzel97,Grosse00}.
The results from the application of the different measures on DNA sequences, 
as well as from analyses making use of several measures together 
(e.g. see \cite{Guharay00}), agree in that there is a different 
structure of the symbolic sequence in the coding and non-coding regions 
and that the non-coding symbolic sequences are more correlated than the 
coding ones (see the review in \cite{Buldyrev98} and references therein).
These findings are certainly useful, but they are based on the prior
identification of the coding and non-coding regions.
  
Coding regions are organized in clusters called genes, which contain 
also non-coding regions, called introns, in-between the coding regions,
called exons.
Thus intergenic regions have pure non-coding character, while genes are
of mixed character since they contain both introns (non-coding)
and exons (coding).
We choose to pursue the analysis at the lower level of knowledge of the
DNA sequence, namely when only the genes are identified and not the 
exons and introns in each gene.
This analysis is more involved because the mechanism underlying
the gene expressions is corrupted by the introns and the objective here 
is to investigate whether the statistical properties for genes and 
intergenic regions can be distinguished. 

The statistical aspects we focus on are, the distribution of homologous
nucleotides (successive identical symbols in the DNA sequence)
\cite{Dokholyan97,Stanley99,Almirantis99,Provata02}, the
general correlation between two nucleotides being apart by a varying 
displacement (quantified by the mutual information of the symbols
being apart) \cite{Ebeling87}, and the prediction of a nucleotide at a 
certain position given the other nucleotides in the sequence (by creating
a new algorithm making use of a local prediction technique modified for 
symbolic time series).            
To facilitate the correct interpretation of the results of this 
analysis, we make the same statistical analysis on artificial symbolic
sequences for which we know the generating mechanism.
We use data from a purely stochastic system, a linear stochastic
system and a simple chaotic system.  
The results of the analysis on genes and intergenic regions combined 
also with the results on the artificial systems give some insight onto the 
level of structure and randomness in the two types of DNA sequences.
In order to assess rigorously the level of randomness or correlations
in the two types of DNA sequences we apply also statistical tests using 
the method of surrogate data \cite{Schreiber99b, Kugiumtzis01a}. 

The paper is organized as follows. 
In Section~\ref{DNAdata}, the DNA symbolic sequences 
are described and the artificial and DNA data sets used in the 
statistical analysis are presented.
In Section~\ref{Methods}, the statistical methods are described and
in Section~\ref{Results} the results of the analysis are presented.
Finally, in Section~\ref{Discussion} the results are discussed, the
main conclusions are drawn and open problems are introduced.

\section{DNA and artificial Symbolic Sequences}
\label{DNAdata} 

The primary structure of DNA consists basically of a nitrogenous base 
of four nucleotides, the two purines, adenine (A) and 
guanine (G), and the two pyrimidines, cytosine (C) and thymine (T).
Thus the DNA sequence can be simply considered as a symbolic sequence 
on the four symbols A,C,G,T.
The genes are specific subsequences of nucleotides encoding information
required to construct proteins.
In primary organisms it is found that the genes cover a large proportion
of the whole DNA. 
In higher organisms, such as the human, genes are more sparsely distributed
and the coding sequences constitute a very small proportion of the genome.
Thus, in terms of data availability, it is difficult to pursue a 
statistical analysis on the two distinct types of the DNA
sequence (the genes and intergenic regions) on higher organisms.
In our analysis we use a large segment of the Chromosome 1 of 
{\em Arabidopsis Thaliana} (a plant), which we denote CAT1.
The genes in CAT1 have been identified and the proportion of genes 
is about 50\%, relatively high for higher eucaryotes.
The complete sequence of CAT1 has 29640317 bases, but we use a 
smaller segment (from base 3760 to base 1000232) to make two sequences, 
one joining together the genes (of total length 574395) and another joining 
together the intergenic regions (of total length 425837).
The median size for genes and for intergenic regions is 1608 and 1170,
respectively, and the first and third quartiles are (1045, 2542) and
(687, 2030), respectively.
 
In order to interpret the results of the statistical analysis
we use also artificial symbolic sequences on four symbols from 
well-known systems. 
The simulated symbolic sequences are designed to possess the probability 
distribution of the four symbols of the DNA symbolic sequence.
This means that to each of the gene and intergenic sequence corresponds 
a different symbolic sequence derived from the same system. 
The systems we consider here generate numeric time series and we transform 
them to symbolic.
We choose: a purely stochastic system (a), a linear stochastic system
with strong and weak correlations (b1 and b2) and a chaotic system (c). 
The respective toy models are: (a) a purely white noise (uniform in [0,1]); 
(b1) a first order autoregressive model, AR(1), with coefficient $\phi=0.9$ 
to assign for strong autocorrelations and (b2) $\phi=0.4$ to assign for 
weak autocorrelations; (c) the logistic map $s_{i+1} = 4 s_i(1-s_i)$ at 
chaotic regime.

The transformation of the simulated numeric time series to symbolic
time series is done with respect to the probability distribution of the 
A,C,G,T symbols of the DNA sequence.
For this we select 3 breakpoints in the range of the data of the
numeric time series, so that the relative frequency in each of the 4 bins 
(formed by the minimum data point, the 3 breakpoints and the maximum data 
point) is equal to the relative frequency of an assigned symbol of the DNA 
sequence.
Let $\{x_i\}_1^n$ be the DNA symbolic sequence, where 
$x_i \in \{\mbox{A,C,G,T}\}$, and $\{y_i\}_1^n$ be the artificial 
symbolic sequence derived from a numeric time series 
($y_i \in \{\mbox{A,C,G,T}\}$).
Then by construction of $\{y_i\}_1^n$ we have
\[
p_x(a) = p_y(a), \quad \mbox{or} \quad p(x=a) = p(y=a) \quad 
\forall \,\, a=\mbox{A,C,G,T}, 
\]
where $p_x$ is the probability mass function of $x$ and $p_x(a)=p(x=a)$ 
is the probability that $x=a$.  
Note that to each of the two types of DNA sequences corresponds a
different symbolic sequence derived from the same artificial numeric
time series by a classification mechanism that is defined by the 
base probabilities of the corresponding DNA sequence. 
  
\section{Methods}
\label{Methods} 

A full statistical description of a symbolic sequence 
would require the estimation of the $w$--joint probability function 
$p_x(x_i,x_{i-1},\ldots,x_{i-w+1})$ for sufficiently large window $w$.
For limited size sequences $\{x_i\}_1^n$, reliable estimations can be 
achieved only for very small $w$.
Therefore, we rely instead on statistical measures related to certain 
aspects of $p_x(x_i,x_{i-1},\ldots,x_{i-w+1})$.
In the following, we present the methods of our statistical analysis 
including a hypothesis test that makes use of surrogate data.

\subsection{Probability distribution of tuples of symbols}

As a first statistical approach towards the investigation of
correlations in symbolic sequences we consider the probability 
distribution of the size of clusters of an identical symbol.
Note that if the symbolic sequence $\{x\}$ is completely independent 
then for a cluster of size $m$ it should be
\be
p_x(a^m) = p_x(\underbrace{aa\cdots a}_{m \,\, \mbox{\small times}}) = 
p_x(a)^m, 
\quad \mbox{where} \quad a=\mbox{A,C,G,T},
\label{eq:cluster}
\ee
where $p_x(a^m)$ is the probability that symbol $a$ occurs $m$ times
sequentially and $a^m$ is called the $m$--tuple of $a$.
The probability $p_x(a^m)$ is estimated by the relative frequency of 
the occurrence of the $m$-tuples of $a$ in the symbolic sequence.
For a tuple of length $w$, $p_x(a^w)$ is actually the evaluation 
of $p_x(x_i,x_{i-1},\ldots,x_{i-w+1})$ in the case 
$x_i=x_{i-1}=\cdots=x_{i-w+1}=a$.

\subsection{Mutual information}

The mutual information $I(x,y)$ of two variables $x$ and $y$ measures the 
general correlation between $x$ and $y$ and it is defined as 
\cite{Shannon49,Fraser86}
\[
I(x,y) = \sum_{a,b} p_{xy}(a,b) \log \frac{p_{xy}(a,b)}{p_x(a)p_y(b)},
\] 
where $a$ and $b$ in the double sum are the possible values $x$ and $y$ 
can take.
For time series or spatial sequences the mutual information regards the 
variables that are apart by a lag or displacement $\tau$,
namely $x_i$ and $x_{i-\tau}$, and the mutual information is then denoted 
$I(\tau)$. 
For the symbolic sequences considered here there are 4 distinct
base probabilities, i.e., $p_{x_i}(a)=p_{x_{i-\tau}}(a)$ for 
$a=\mbox{A,C,G,T}$, and 16 joint probabilities for each $\tau$, 
i.e., $p_{x_i x_{i-\tau}}(a,b)$ for $a,b=\mbox{A,C,G,T}$.
The base and joint probabilities are again estimated by the relative 
frequencies computed on the symbolic sequence.
Note that setting $\tau=w-1$, $I(w-1)$ is derived from $p_x(x_i,x_{i-w+1})$, 
which is the projection of the $w$--joint probability function 
$p_x(x_i,x_{i-1},\ldots,x_{i-w+1})$ onto the first and last component
of the window $(x_i,x_{i-1},\ldots,x_{i-w+1})$. 

\subsection{Identical Neighbor Fit}

The identical neighbor fit is actually a modification of the 
method of nearest neighbor prediction used for nonlinear prediction 
and modeling of numeric time series \cite{Kantz97}.
In the context of symbolic sequences, the prediction problem is to estimate 
a symbol $T$ positions forward when we know the symbols up to the 
current position $i$.
However, for DNA sequences we consider the modeling or fitting problem
rather than the prediction problem since we know all symbols.
For each position $i$ in the symbolic sequence, we want to estimate 
the probability of correct identification of the symbol in position $i+T$
using all other symbols in the sequence, i.e. symbols in positions
$1,\ldots,i,i+T+1,\ldots,n$, where $n$ is the length of the sequence.
The level of fit as defined by this probability constitutes another 
measure of the degree of correlations in the symbolic sequence.  

Similarly to the state space reconstruction of numeric time series,
we assign for each symbol at position $i$ in the symbolic sequence,
the segment $i$ of size $m$ comprised of the $m$ last symbols,
i.e. the symbols in positions $i-m+1,\ldots,i$.  
The parameter $m$ acts here as the embedding dimension for numeric 
time series. 
Thus for each symbol $x_i$, $i=m,m+1,\ldots,n-T$ we assign the
symbolic vector $\mathbf{x}_i = [x_{i-m+1},\ldots,x_i]'$.  
If the working position for the fit is $i$, the target vector is 
$\mathbf{x}_i$, and we want to estimate the symbol $x_{i+T}$.
To do this, we search across the sequence
$\{\mathbf{x}_m,\mathbf{x}_{m+1},\ldots,\mathbf{x}_{i-1},
\mathbf{x}_{i+m+T},\ldots,\mathbf{x}_{n-T}\}$ (we exclude vectors that 
have as components any of $x_{i},\ldots,x_{i+T}$) to find symbolic 
vectors that are identical to $\mathbf{x}_i$, which we call 
{\em identical neighbors} (similarly to the nearest neighbors for 
numeric time series \cite{Kantz97}).
Let us suppose that we found $K$ identical neighbors of $\mathbf{x}_i$ 
in positions $i_1,\ldots,i_K$ and the respective symbols $T$ steps 
forward are $x_{i_1+T},\ldots,x_{i_K+T}$.   
Since we know the actual symbol $x_{i+T}$ we define the prediction
error based on the $k$-th identical neighbor as
\[
e_{i_k}(T) = \left \{ \begin{array}{r@{\quad\mbox{if}\quad}l}
0 & x_{i+T} = x_{i_k+T} \\
1 & x_{i+T} \neq x_{i_k+T}
\end{array} \right.
\]
The error in the prediction of $x_{i+T}$ using all $K$ identical 
neighbors of $\mathbf{x}_i$ is the proportion of {\em false identical 
neighbor predictions} defined as
\[
E_{i}(T) = \frac{1}{K} \sum_{k=1}^K e_{i_k}(T).
\]
If $E_i = 0$ the prediction of $x_{i+T}$ is perfect while if $E_i = 1$ 
the prediction fails completely.

Finally, we average the individual prediction errors over all symbolic 
vectors ($i=m,m+1,\ldots,n-T$) to get a measure of the 
{\em mean identical neighbor error} (MINE) for the whole sequence, 
defined as 
\be
\mbox{MINE}(T) = \frac{1}{n-m-T+1} \sum_{i=m}^{n-T} E_i(T).
\label{eq:mine}
\ee
One can also define the weighted MINE (WMINE) from the weighted average 
with respect to the number $K_i$ of identical neighbors found at each 
individual prediction $i$,
\be
\mbox{WMINE}(T) = \frac{\sum_{i=m}^{n-T} K_i E_i(T)}{\sum_{i=m}^{n-T} K_i}.
\label{eq:WMINE}
\ee 

For a symbolic sequence $\{x\}_1^n$, the maximum prediction 
error level depends on the base probability distribution $p_x(a)$ for each
symbol $a$ and it is reached when the symbolic sequence is
purely random.
It is straightforward to find that the maximum error is   
\be
\max \mbox{MINE}(T) = \max \mbox{WMINE}(T) = 1 - \sum_a p_x(a)^2,  
\label{eq:maxmime}
\ee
where the sum is over the symbols of the sequence.
For a random sequence of 4 equally probable symbols ($p_x(a) = 0.25$ for
every $a$) we get $\max \mbox{MINE}(T) = 0.75$.  

The measure of identical neighbor fit is also related to the joint 
probability function $p_x(x_i,x_{i-1},\ldots,x_{i-w+1})$ through
the conditional probability function 
$p_x(x_{j+T}|x_j,x_{j-1},\ldots,x_{j-m+1})$  
where $w=m+T$ and $i=j+T$. 
The estimation of this conditional probability is actually the problem 
of individual $T$ step ahead prediction for a position $j$ (actually,
the false identical neighbor prediction $E_j(T)$ estimates the 
complementary conditional probability).
In the computation of the total prediction error MINE, we evaluate 
the conditional probability on a subset of the set of all possible  
values $\{x_j,x_{j-1},\ldots,x_{j-m+1}\}$ (those found in the sequence), 
which constitutes a very small fraction when $m$ gets large.    
For $m=1$, the conditional probability reads $p_x(x_{j+T}|x_j)$
and MIME measures essentially the same characteristic as the mutual
information for lag $T$, $I(T)$.
Thus, in terms of information processing, the identical neighbor fit 
can be seen as an extension of mutual information to more than two 
variables, i.e. a measure similar to Shannon-like entropy 
\cite{Ebeling91}. 

\subsection{Hypothesis test with surrogate data}

The working null hypothesis $\mbox{H}_0$ for the DNA sequence is that the 
examined symbolic sequence is completely random, i.e. there are no 
correlations in the sequence.
We generate an ensemble of $M$ surrogate symbolic sequences that 
represent $\mbox{H}_0$.
Each surrogate sequence is simply generated by shuffling the original
sequence. 
Note that in this way, the original base probabilities are preserved in
the surrogate sequence and the sequence is otherwise random.

As discriminating statistic $q$ for the test we consider any of the 
statistics from the statistical methods presented above.
$\mbox{H}_0$ is rejected if the statistic $q_0$ on the original
symbolic sequence does not fall within the empirical distribution of the
statistic $q$ under $\mbox{H}_0$, which is formed by the statistics
$q_1,\ldots,q_M$ computed on the $M$ surrogate symbolic sequences.
A formal test decision can be made using a parametric approach or a 
non-parametric approach (e.g. see \cite{Kugiumtzis01a}).

\section{Results}
\label{Results} 

The statistical analysis with the tools presented in Section\,\ref{Methods} 
extents in two directions: a) investigation in a rigorous 
statistical manner whether the gene sequence and the intergenic sequence
contain significant correlations and assessment of the degree of departure 
from the level of no correlation, and b) comparison of the two 
types of DNA sequences to each other and to some other symbolic sequences 
with known dynamical properties.   

We choose to present the results for each of the three statistical 
methods separately.

\subsection{Results from the probability distribution of symbol clusters}

It is believed that the correlations observed in DNA sequences are mainly due
to non-trivial clustering of homologous nucleotides, which are actually
the successive repetitions of a single symbol in the DNA symbolic sequence. 
Moreover, it is found that the density of clusters of identical symbols 
in non-coding DNA regions is atypically high suggesting long range 
correlations while the density of the same clusters in coding DNA regions 
is lower and consistent to short range correlated symbolic sequences
\cite{Almirantis99, Provata02}. 

\begin{figure}[htb] 
\centerline{\hbox{\includegraphics[width=4.5cm,keepaspectratio]{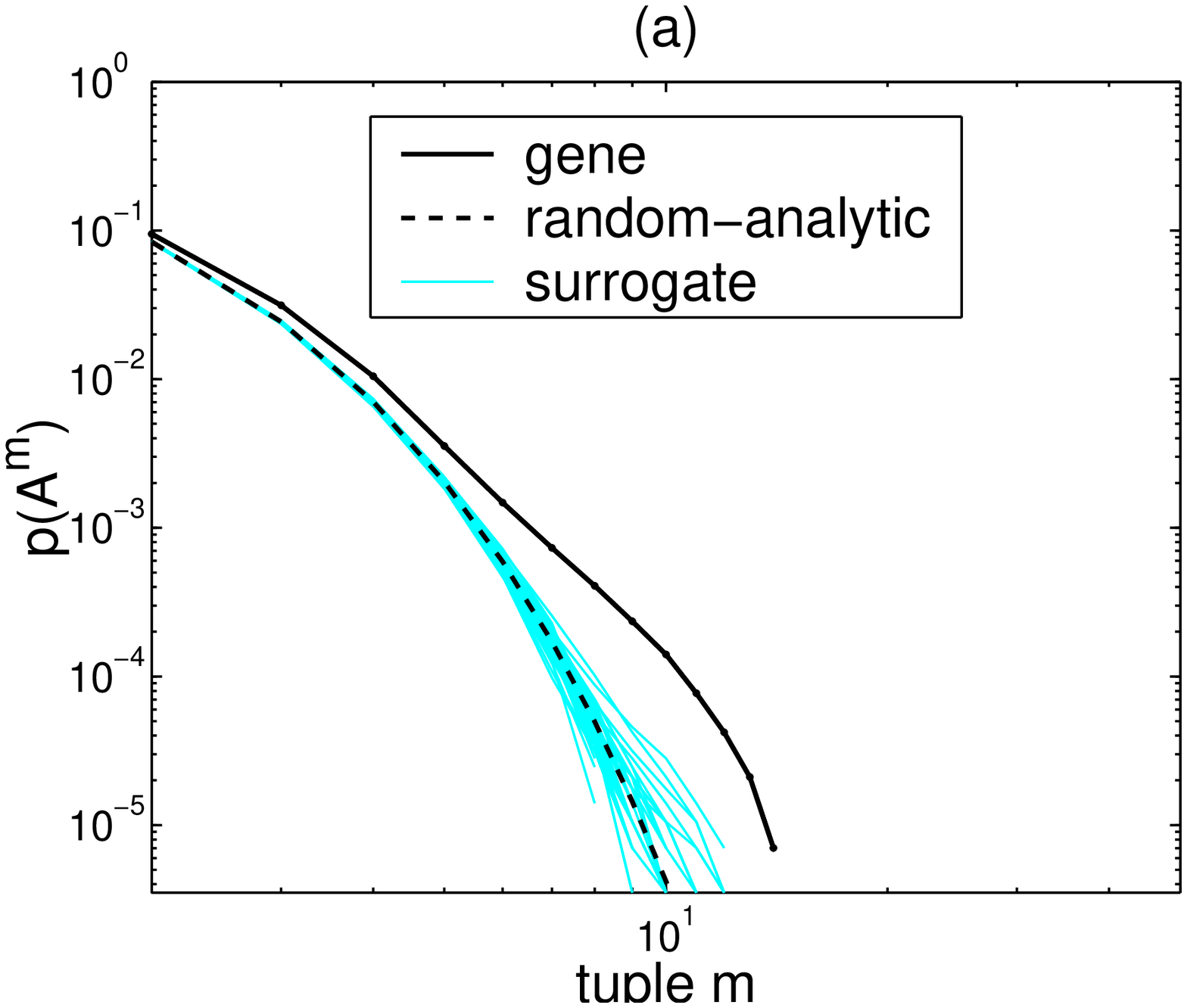}
	\includegraphics[width=4.5cm,keepaspectratio]{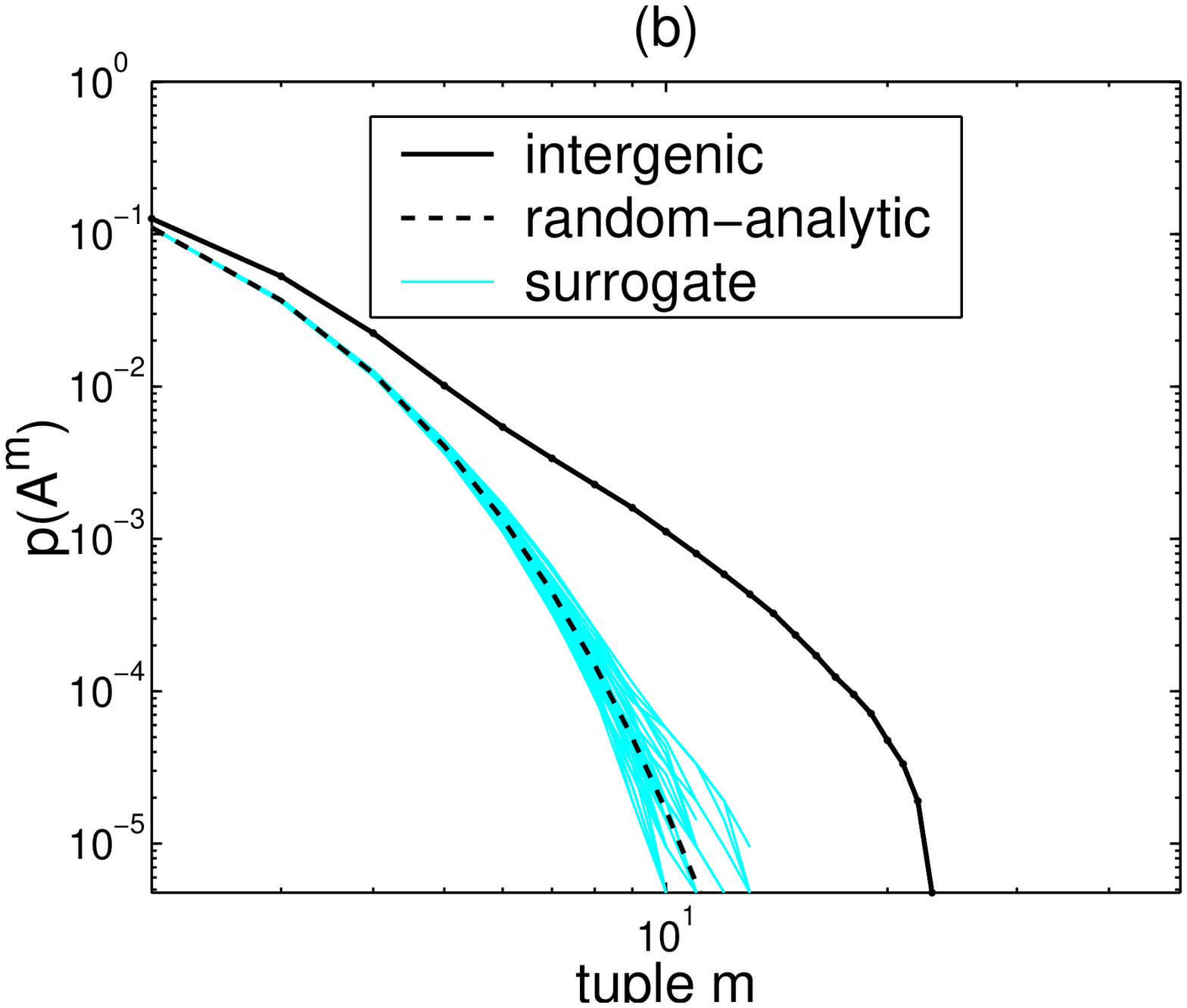}
	\includegraphics[width=4.5cm,keepaspectratio]{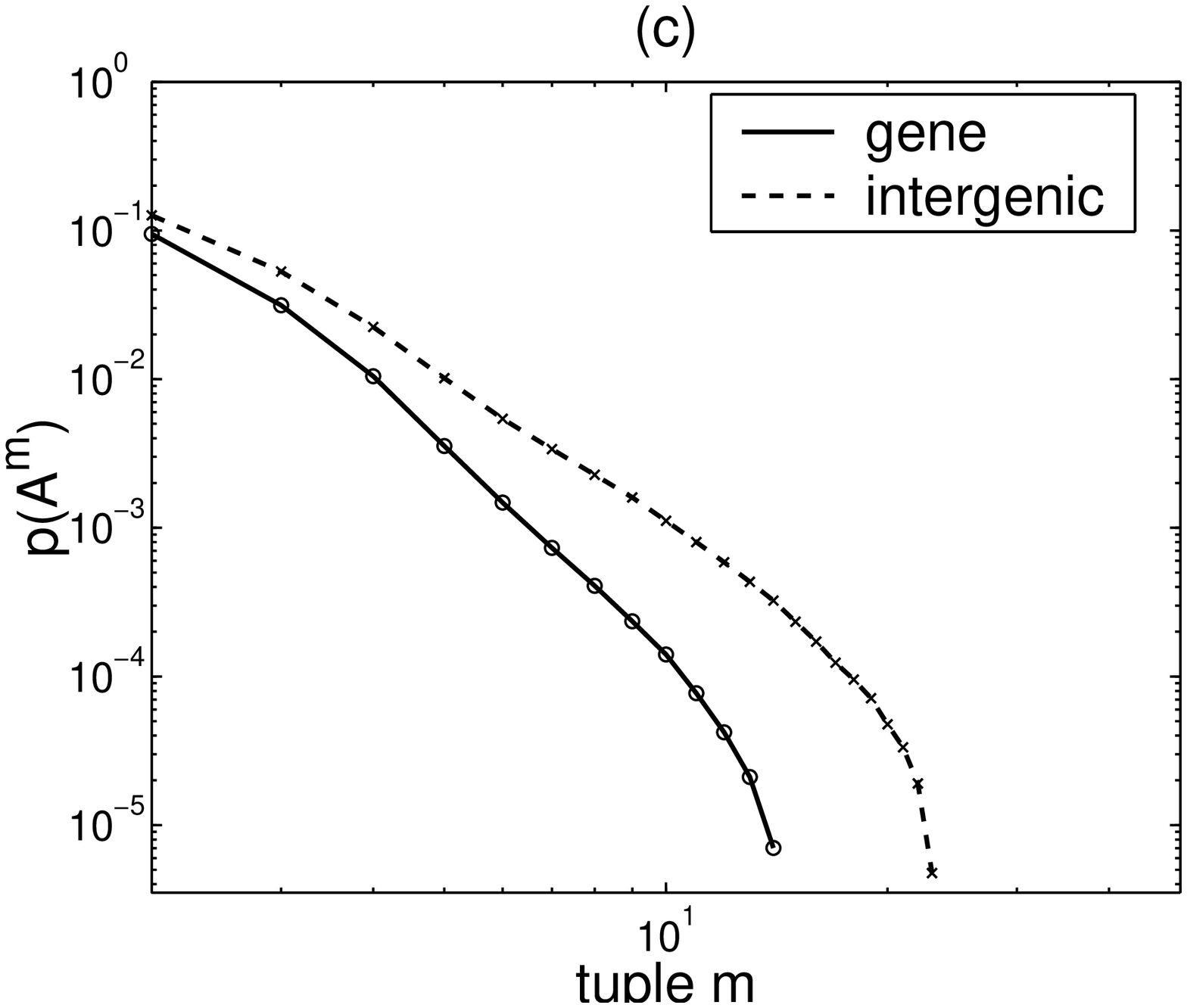}}}
\centerline{\hbox{\includegraphics[width=4.5cm,keepaspectratio]{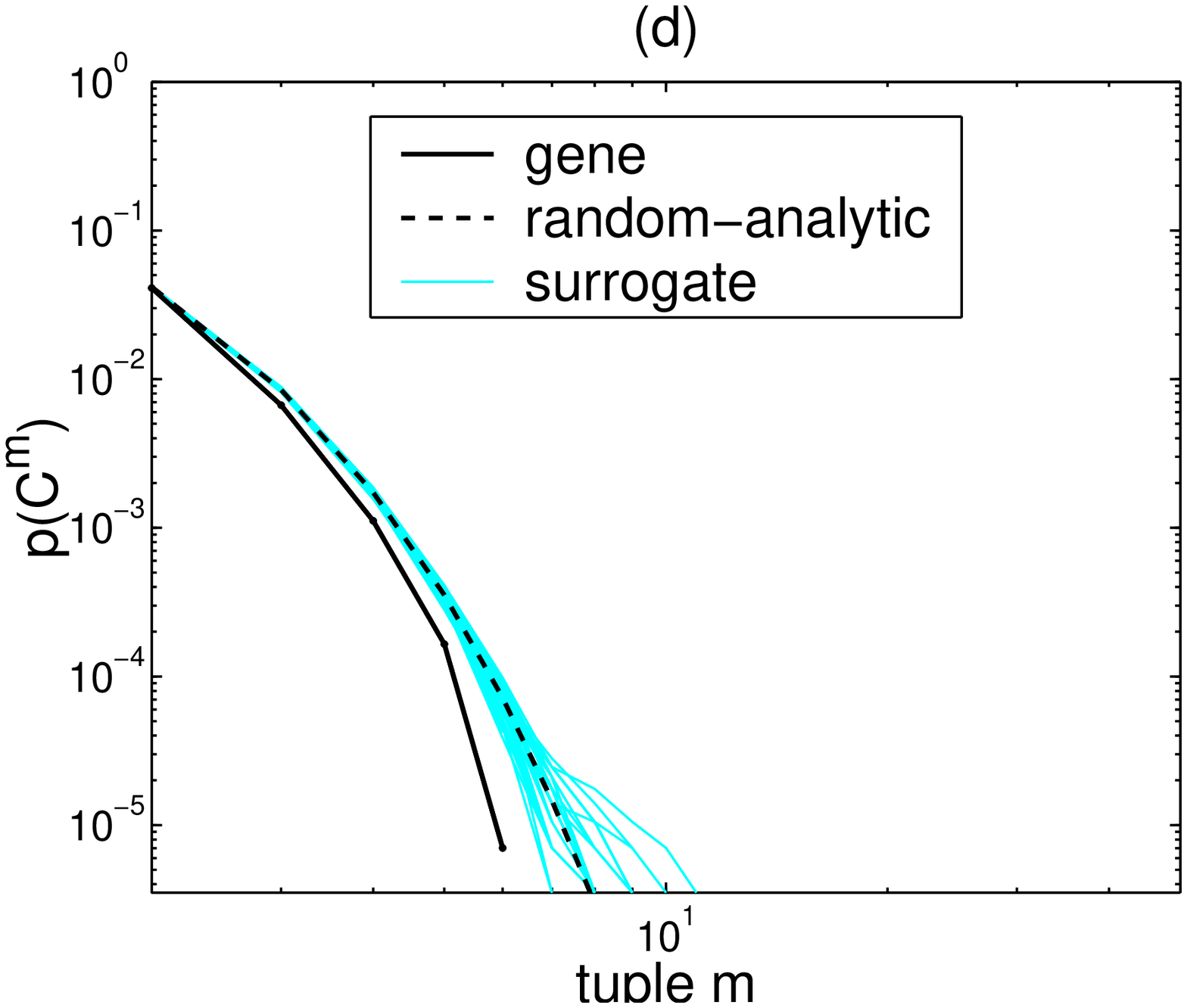}
	\includegraphics[width=4.5cm,keepaspectratio]{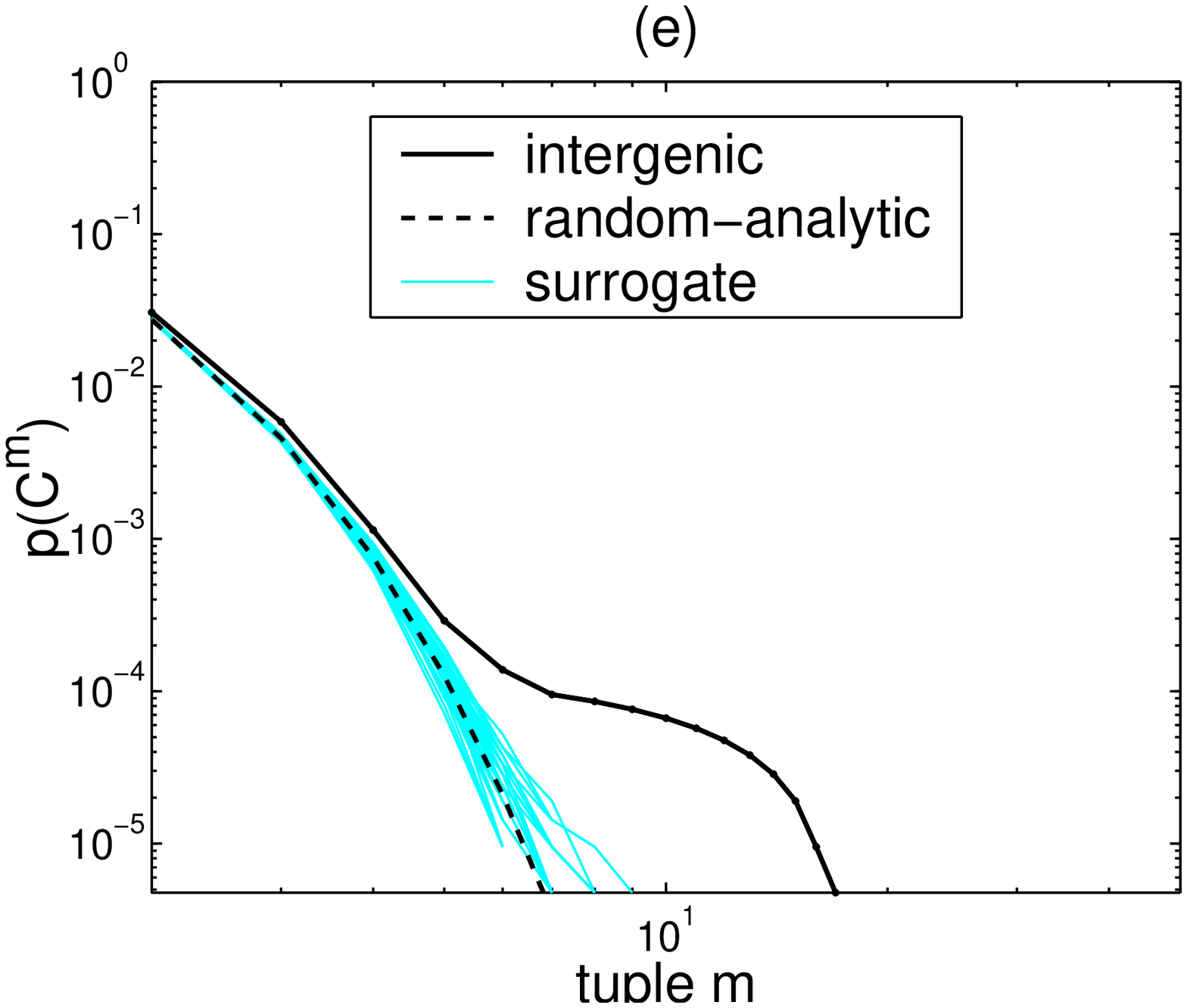}
	\includegraphics[width=4.5cm,keepaspectratio]{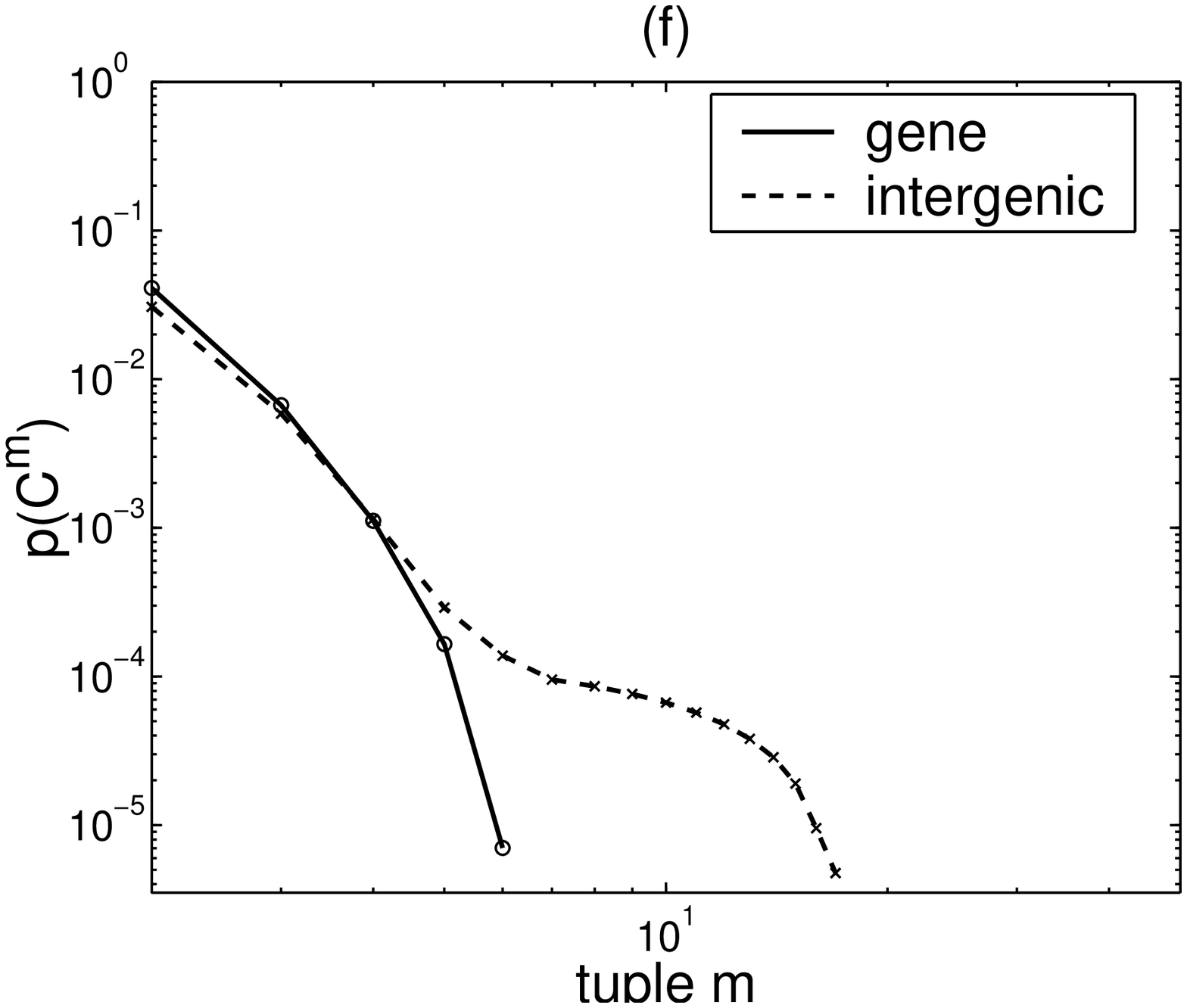}}}
\centerline{\hbox{\includegraphics[width=4.5cm,keepaspectratio]{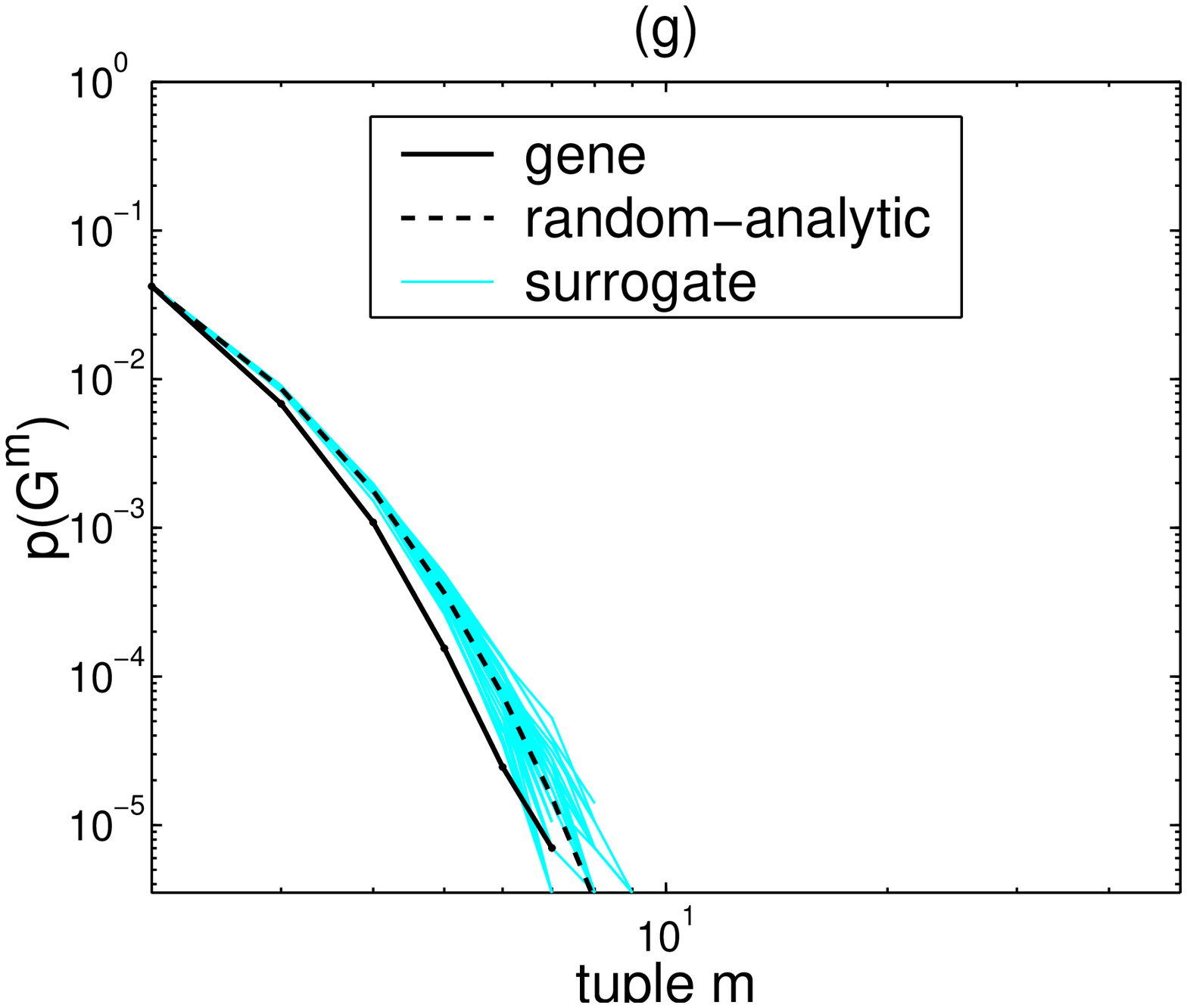}
	\includegraphics[width=4.5cm,keepaspectratio]{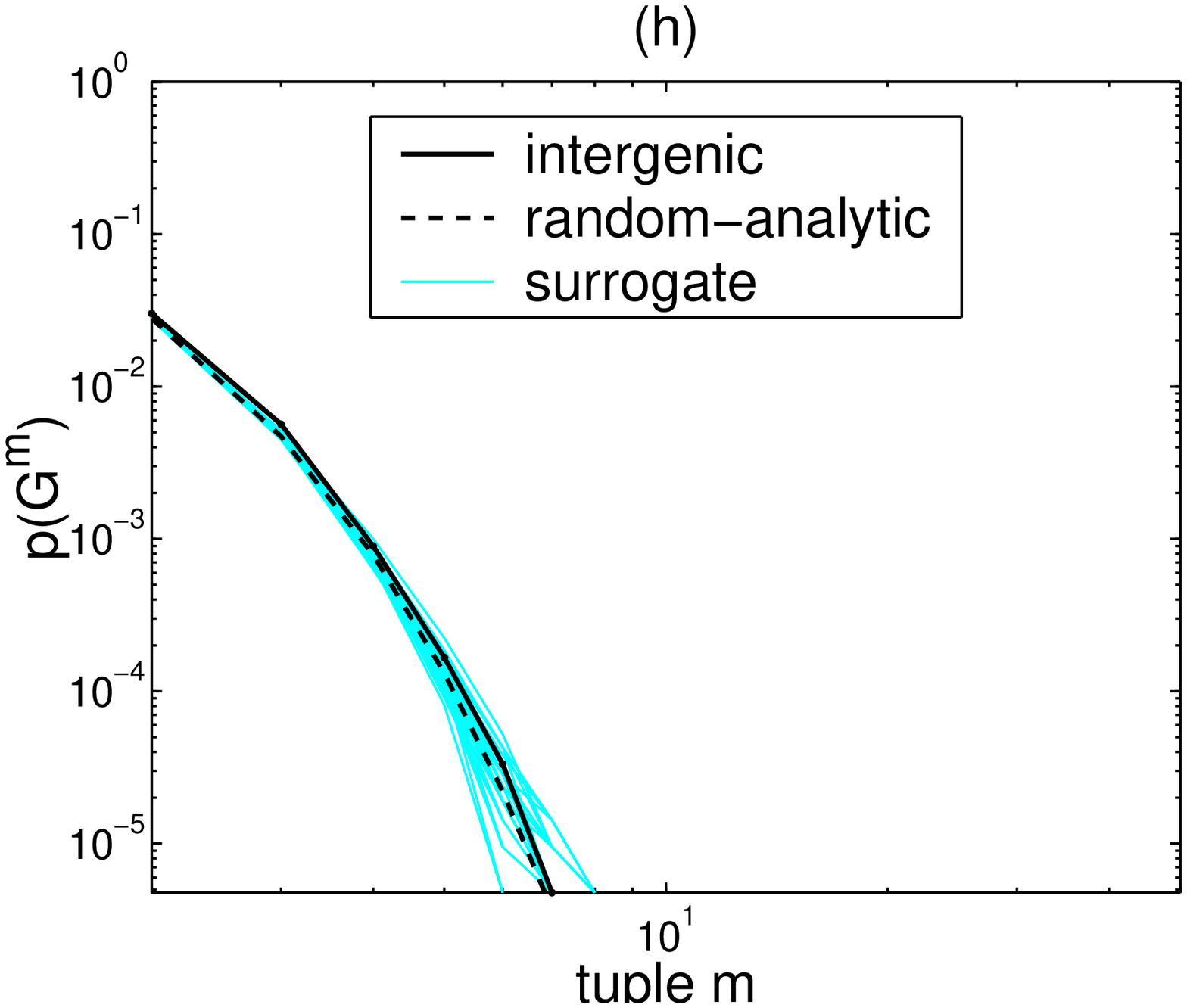}
	\includegraphics[width=4.5cm,keepaspectratio]{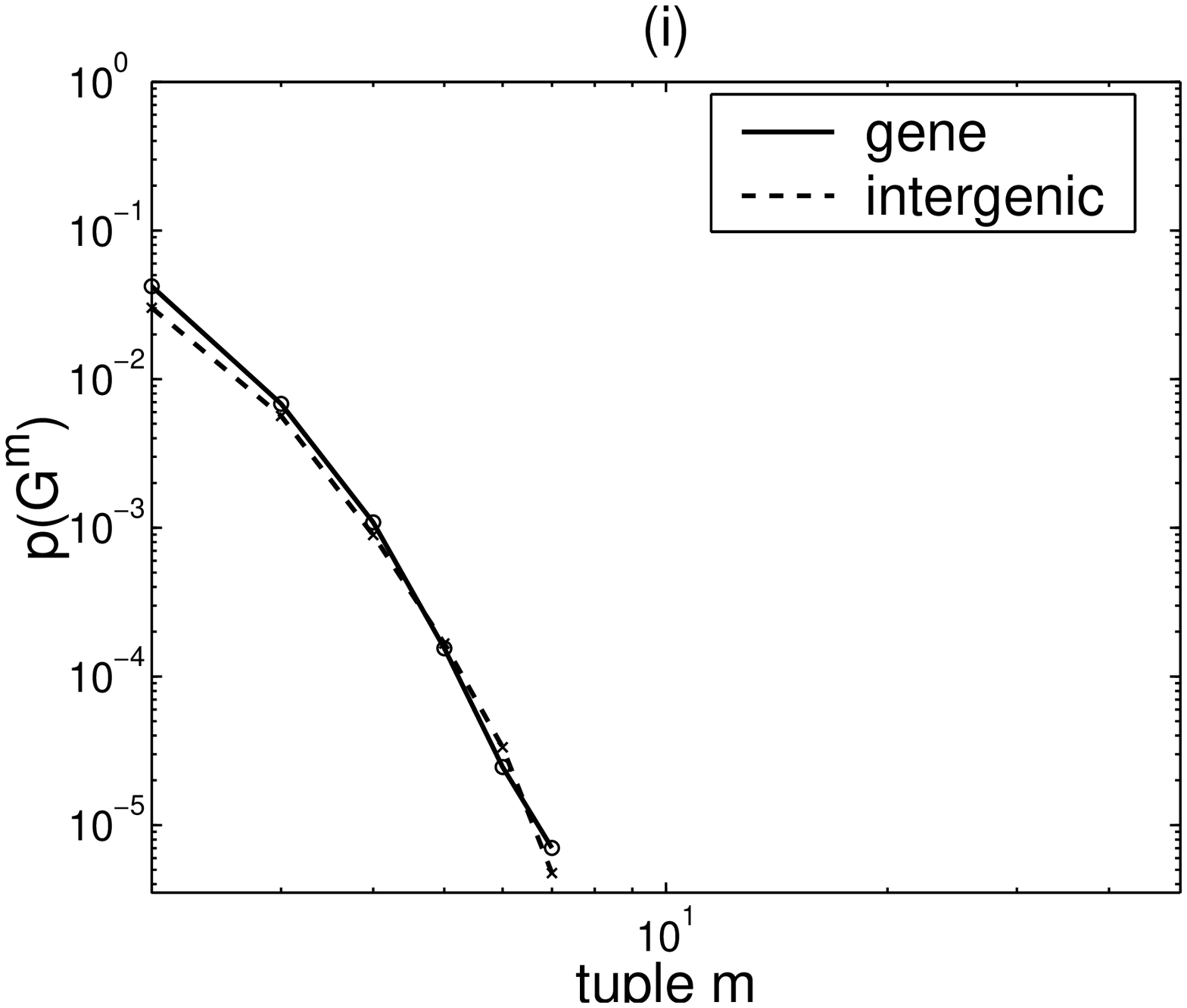}}}
\centerline{\hbox{\includegraphics[width=4.5cm,keepaspectratio]{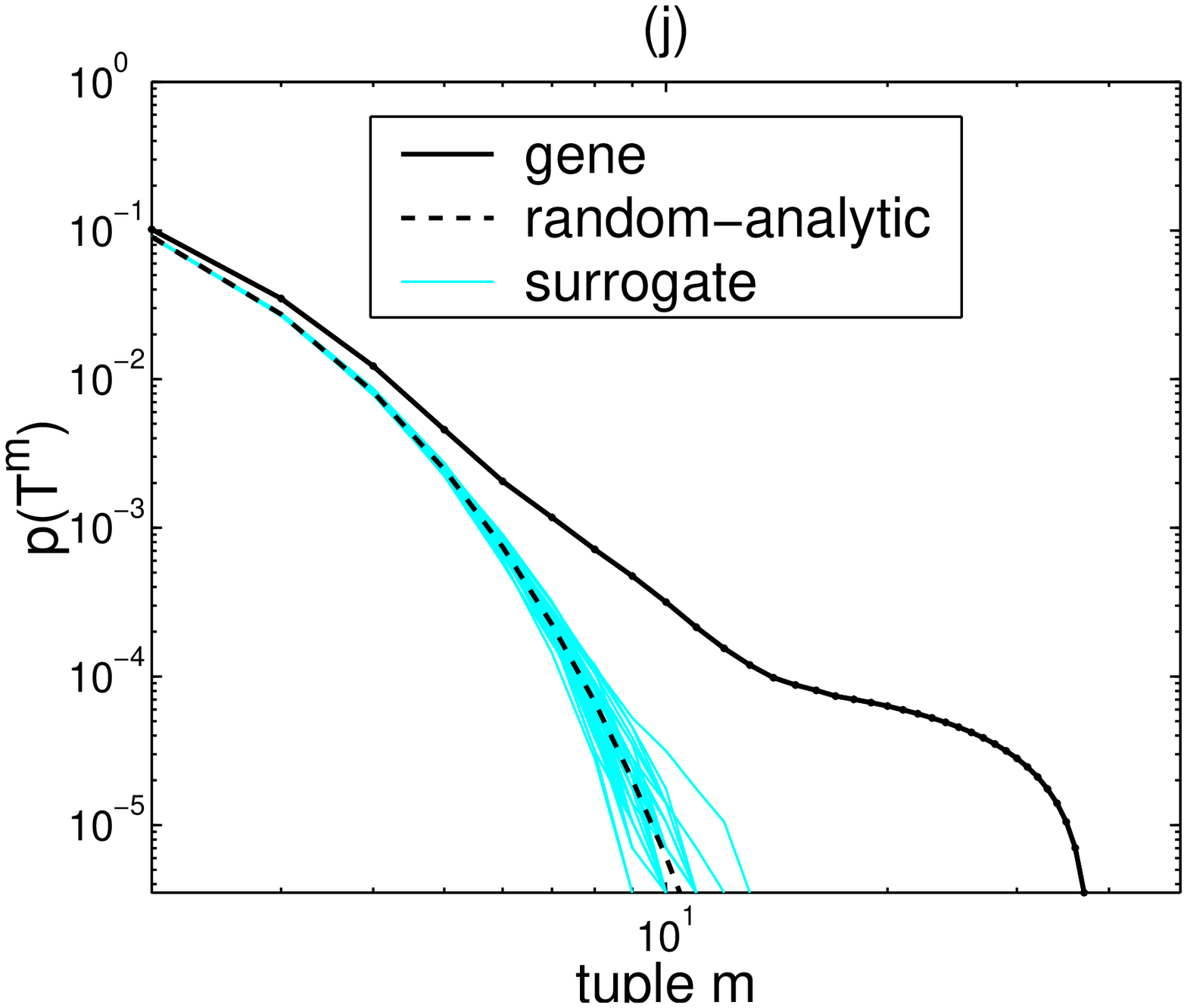}
	\includegraphics[width=4.5cm,keepaspectratio]{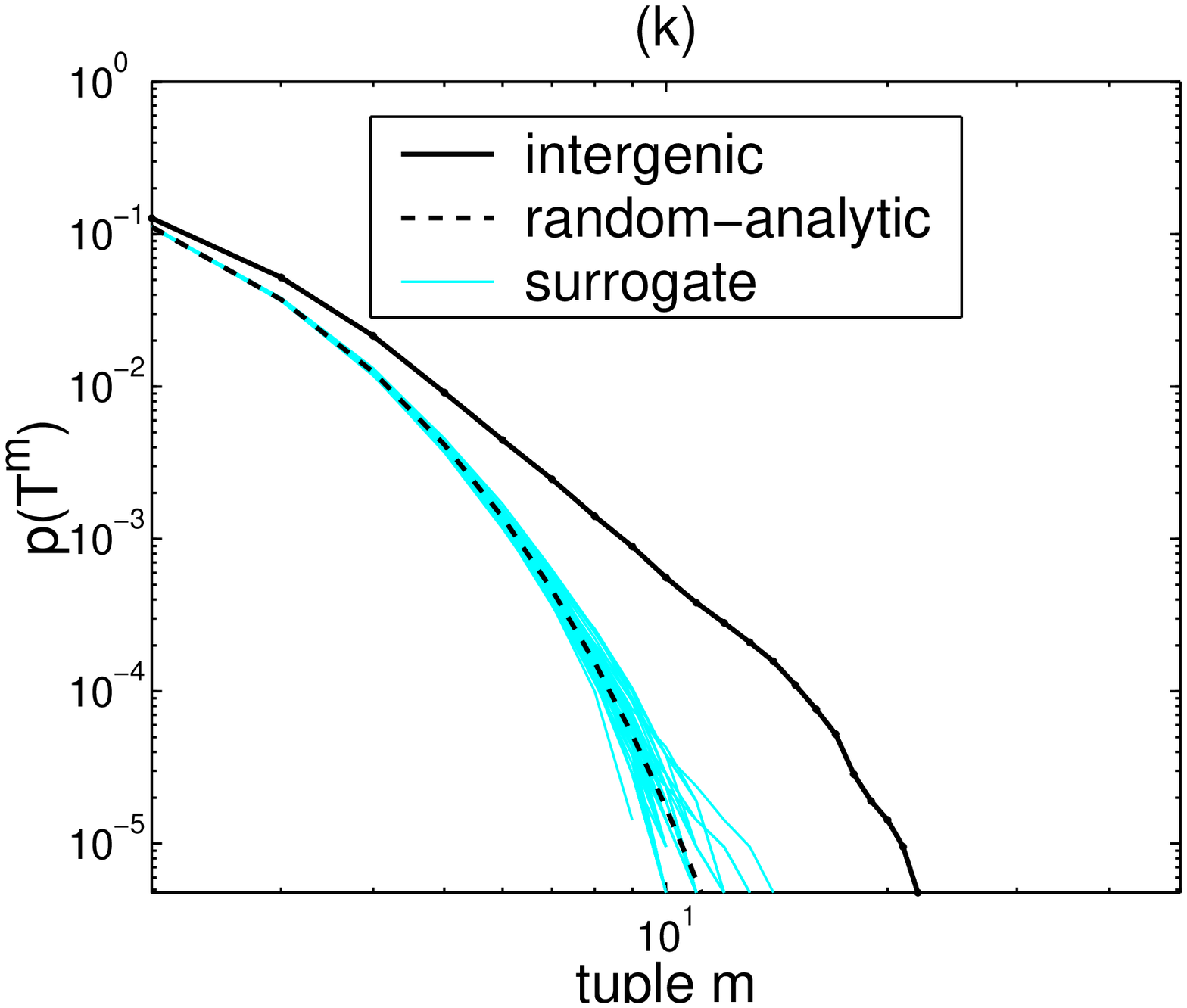}
	\includegraphics[width=4.5cm,keepaspectratio]{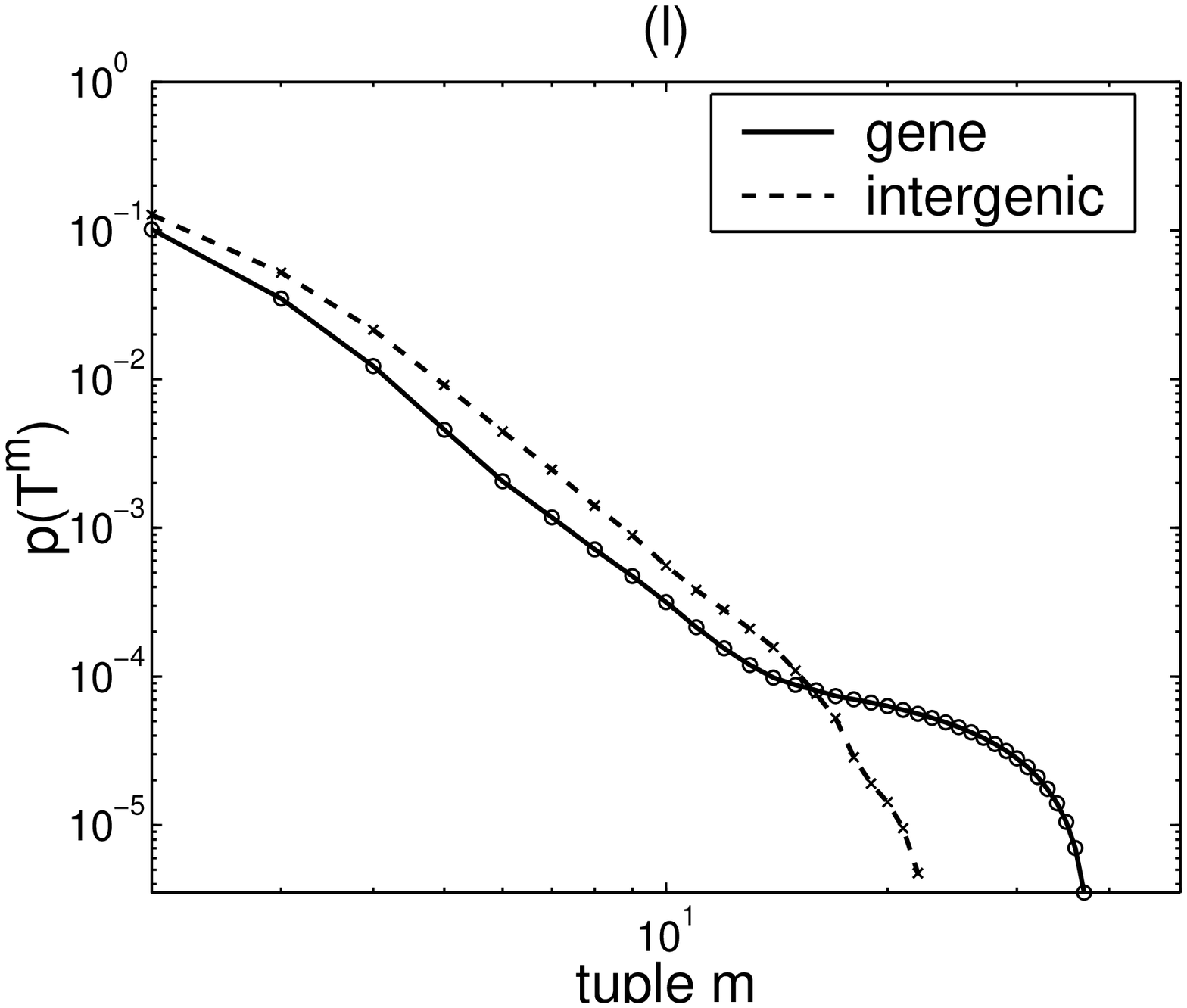}}}
\caption{(a) The graph of the sample probability function of tuples of 
symbol A evaluated for a gene CAT1 sequence of 285000 bases and 40 
surrogate sequences of it.
Superimposed is also the analytic probability function assuming the sequence
is random, as shown in the legend.
(b) The same as in (a) but for an intergenic CAT1 sequence of 210000 bases.
(c) The graphs of the probability functions of tuples of the symbol A 
for the gene sequence in (a) and the intergenic sequence in (b).
The same results as in (a), (b) and (c) are shown for symbol C in panels 
(d), (e), (f), for symbol G in panels (g), (h), (i) and for symbol T in 
panels (j), (k), (l).}
\label{clustersur}
\end{figure}

In a similar way, we study the density of clusters of homologous symbols 
on genes and intergenic sequences.
A gene is a mixture of coding and non-coding parts and according
to the above findings it is expected to contain some form of long range 
correlation due to the non-coding parts in it.
However, the correlation in the genes should be less than the 
correlation in the intergenic sequences which consist only of non-coding 
DNA.
This reasoning is only partially confirmed by the results from the 
probability distribution of symbol clusters.
Figure~\ref{clustersur} shows the probability distribution of $m$-tuples 
of the symbols A, C, G and T for a gene sequence of 285000 bases and 
an intergenic sequence of 210000 bases, both from CAT1.
Superimposed are also the graphs of the same probability function evaluated
for each of the $M=40$ surrogate sequences as well as the analytic 
probability mass function under the assumption of statistical independence.

For both gene and intergenic CAT1 sequences, the  
probability function of tuples of the symbols A and T 
(in Fig.~\ref{clustersur}a,b,j and k) is distinctly higher from  
the analytic probability function under the assumption of complete 
randomness.  
This is confirmed by the surrogate data as their probability distribution 
of tuples is always concentrated along the theoretical probability function.
Obviously, we can reject $\mbox{H}_0$ that the original DNA 
sequence (gene and intergenic regions) has no correlations at high confidence 
levels taken as statistics the relative frequency of any $m$-tuple of 
A or T.  
For symbols C and G the probability of $m$-tuples for the gene DNA sequences
are slightly lower than the analytic probabilities and the probabilities for
the surrogates but the differences are statistically significant for a long
range of $m$-tuples (see Fig.~\ref{clustersur}d and g).
For the intergenic DNA, this difference vanishes for C whereas for G remarkably 
large tuples of $C$ occur with non-zero probability (see Fig.~\ref{clustersur}e
and h).  

For symbols C and G the distribution of the tuples for the DNA sequences
does not differ significantly from random apart from the case of symbol $C$ 
and intergenic sequence, where remarkably large tuples of $C$ occur with 
non-zero probability (see Fig.~\ref{clustersur}e).

From the results in Fig.~\ref{clustersur} it cannot be universally 
concluded that the probability distribution of the tuples of the 
symbols deviate from randomness more for the intergenic sequence 
than for the gene sequence. 
Actually, it seems to hold only for symbol A, as shown in 
Fig.~\ref{clustersur}c.
This is expected because long repetitions of A have been observed in 
higher organisms.
The opposite effect is observed for large tuples of $T$
(see Fig.~\ref{clustersur}l). 

\begin{figure}[htb] 
\centerline{\hbox{\includegraphics[width=4.5cm,keepaspectratio]{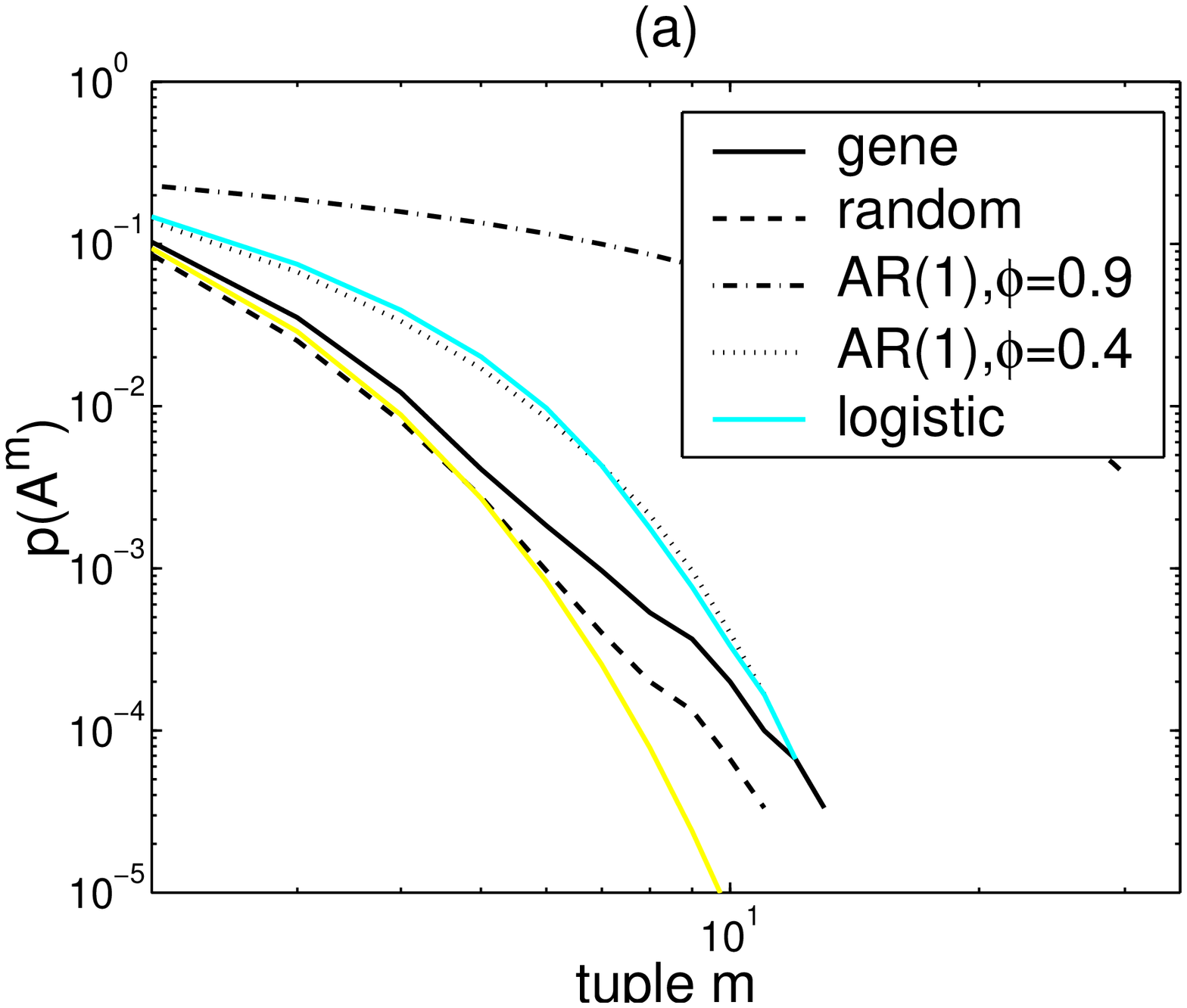}
	\includegraphics[width=4.5cm,keepaspectratio]{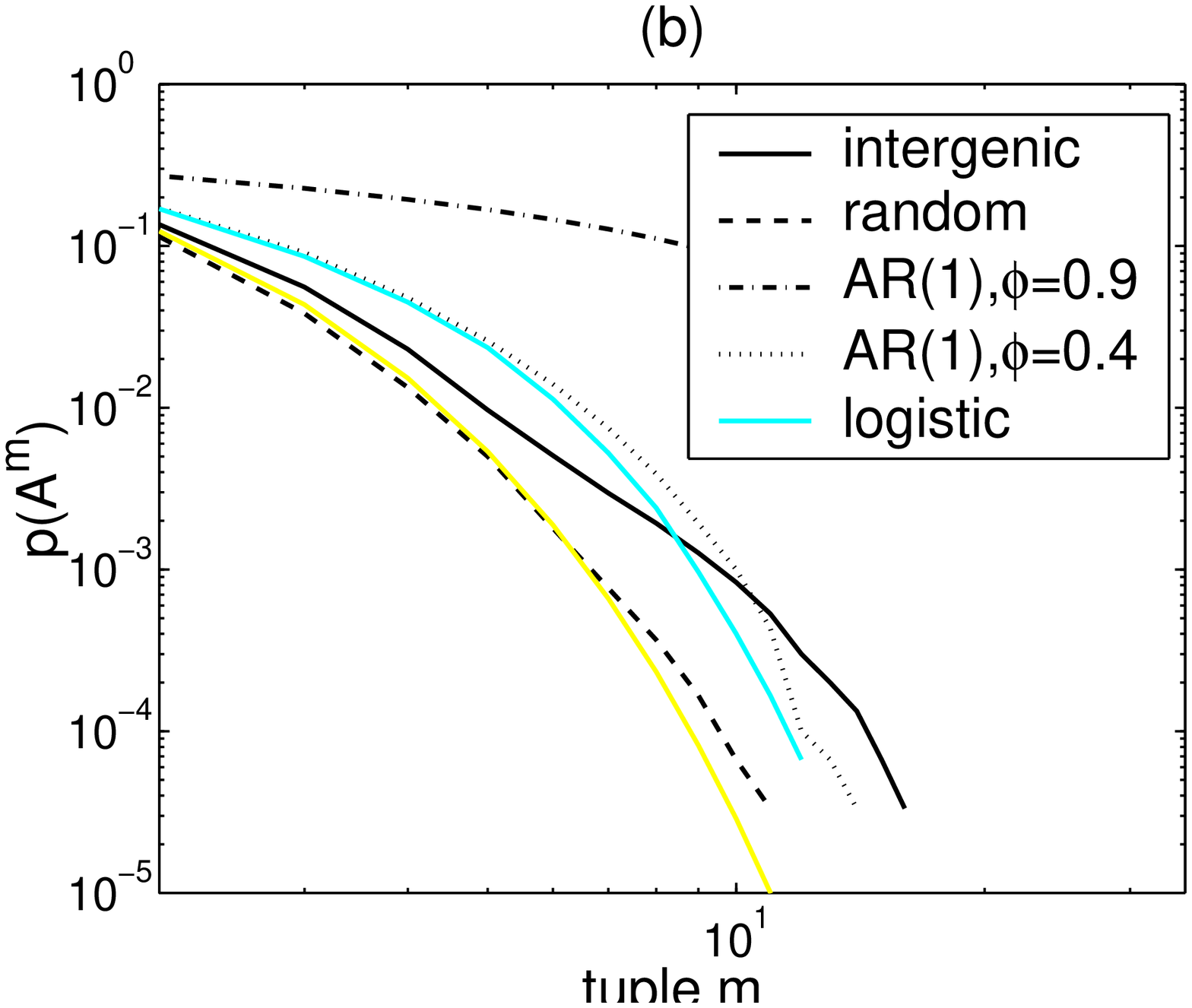}}}
\centerline{\hbox{\includegraphics[width=4.5cm,keepaspectratio]{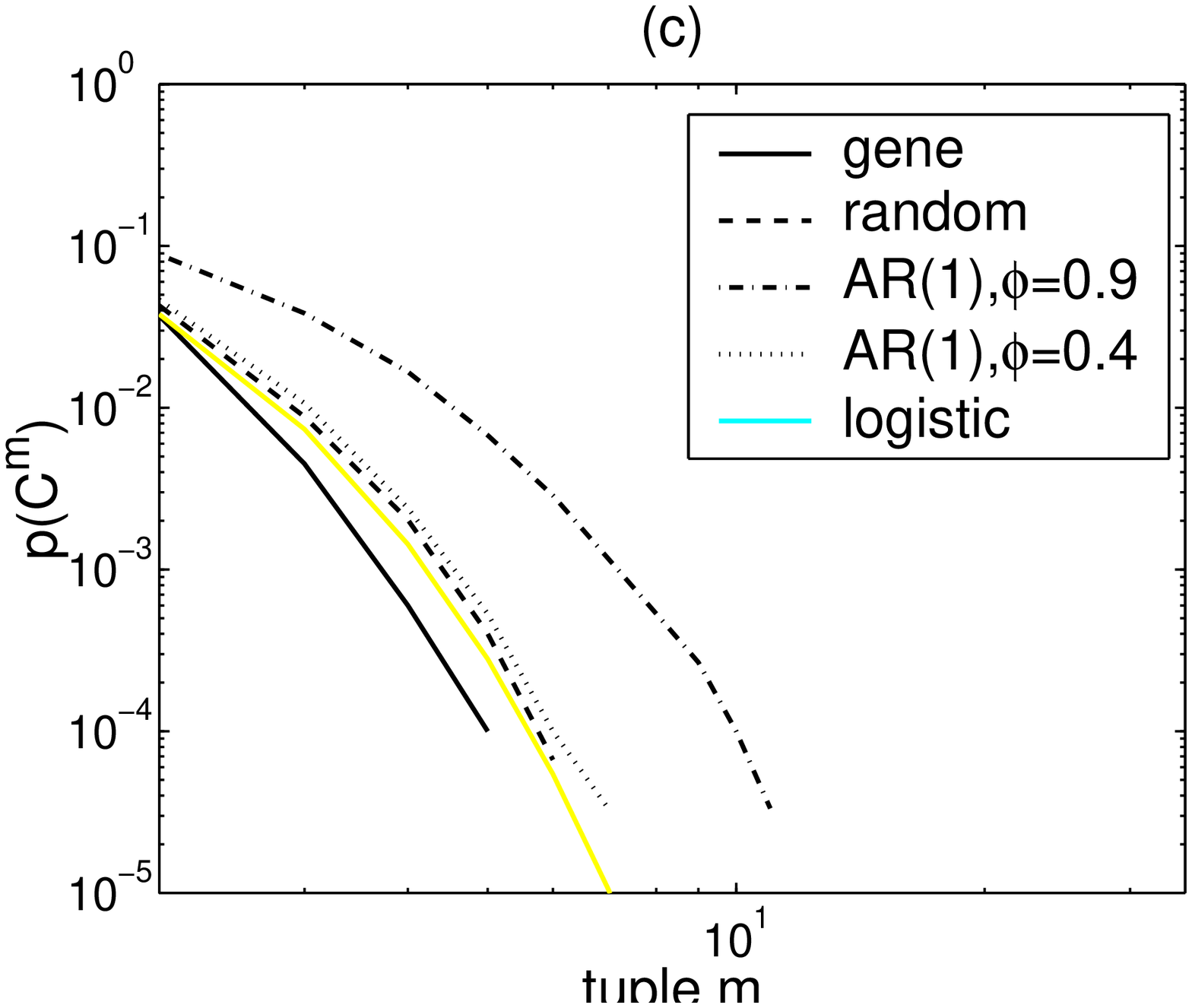}
	\includegraphics[width=4.5cm,keepaspectratio]{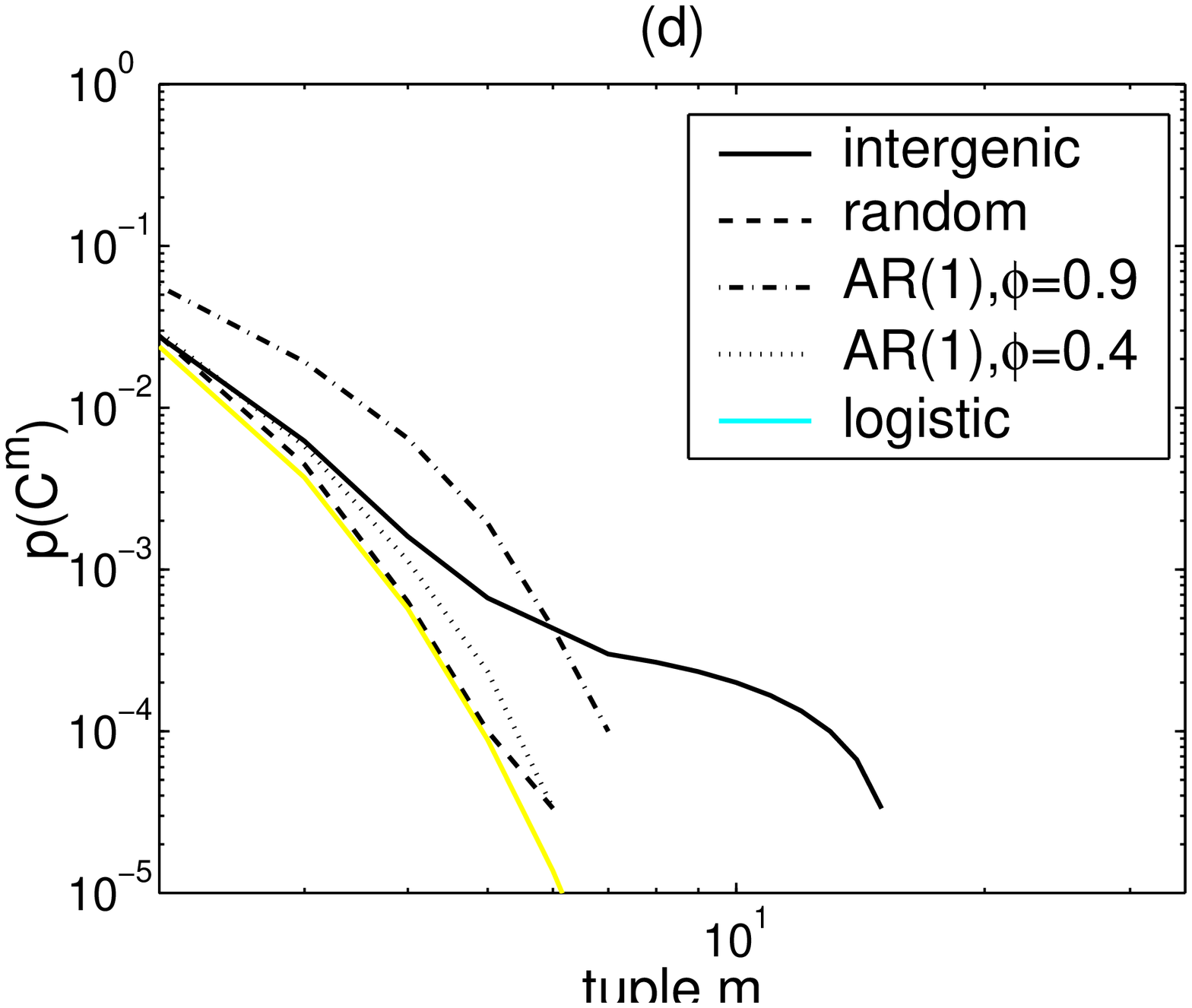}}}
\centerline{\hbox{\includegraphics[width=4.5cm,keepaspectratio]{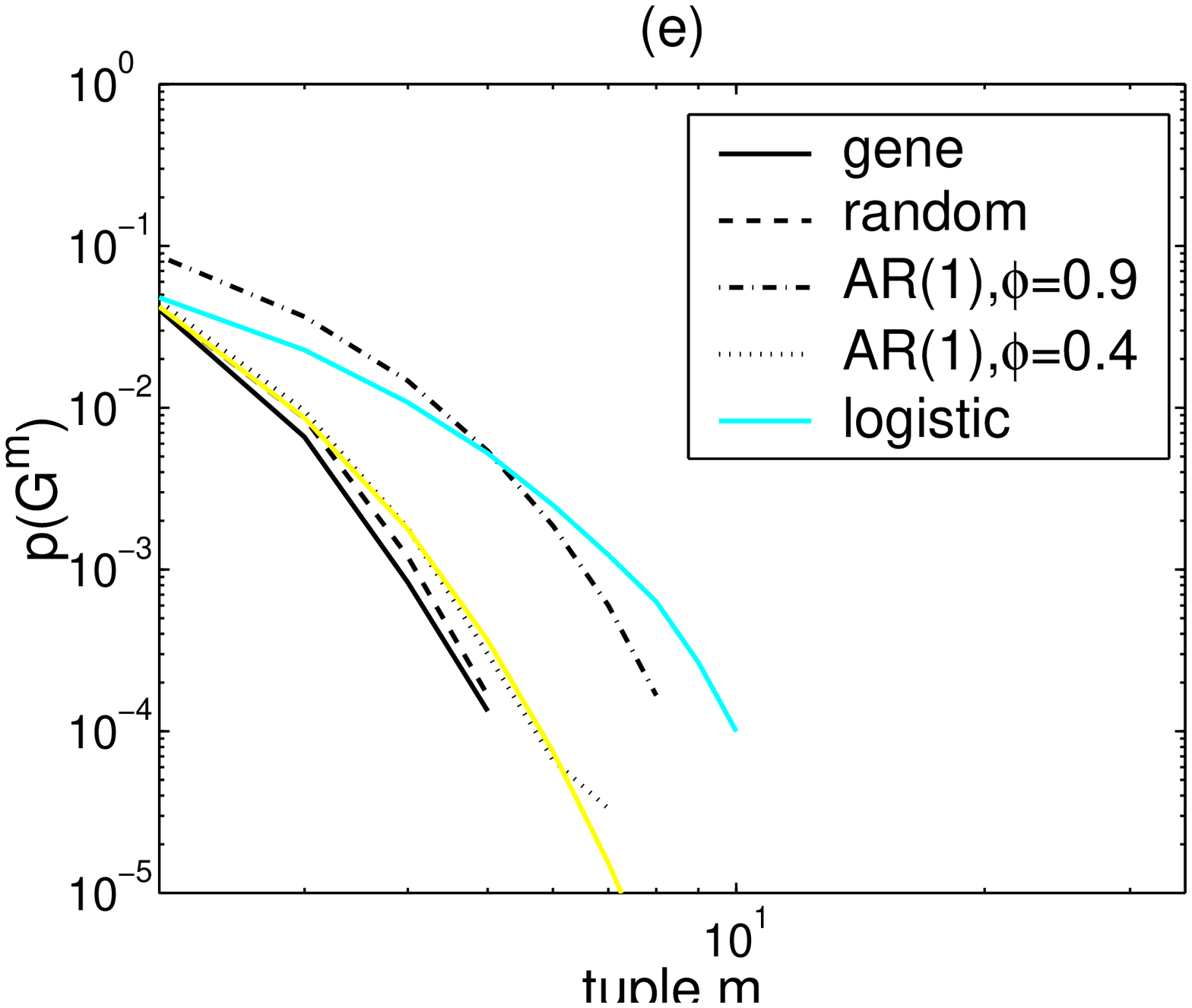}
	\includegraphics[width=4.5cm,keepaspectratio]{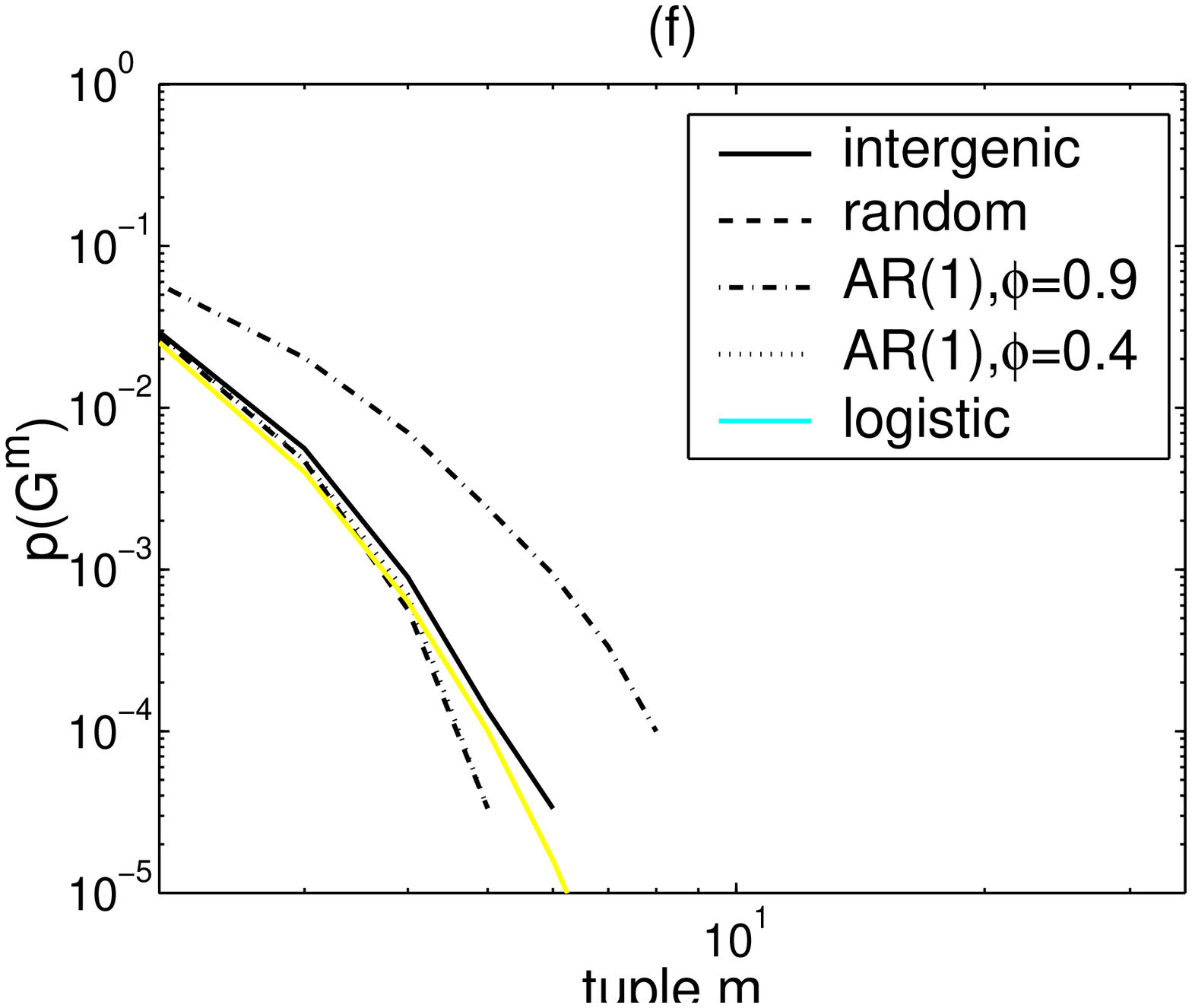}}}
\centerline{\hbox{\includegraphics[width=4.5cm,keepaspectratio]{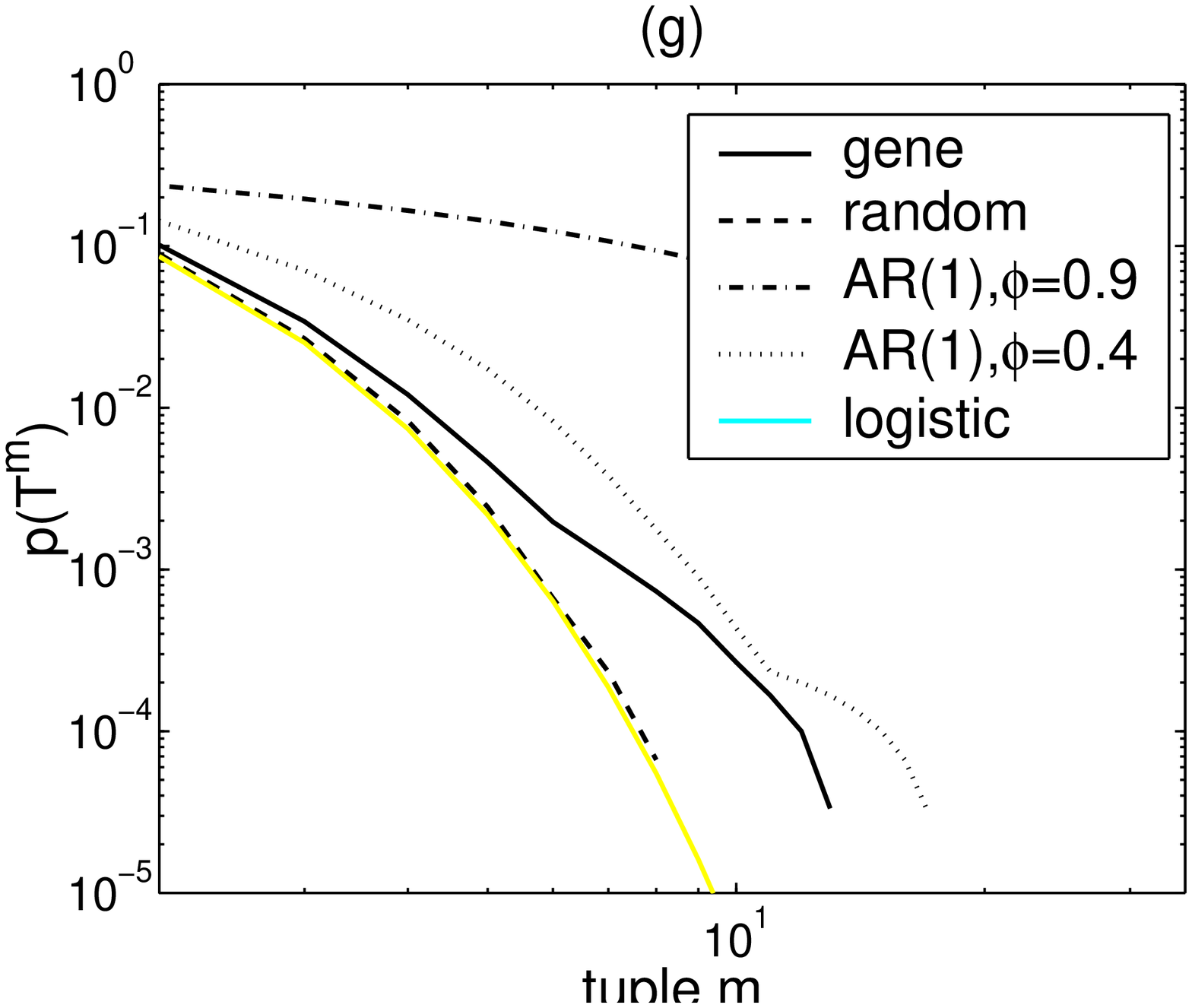}
	\includegraphics[width=4.5cm,keepaspectratio]{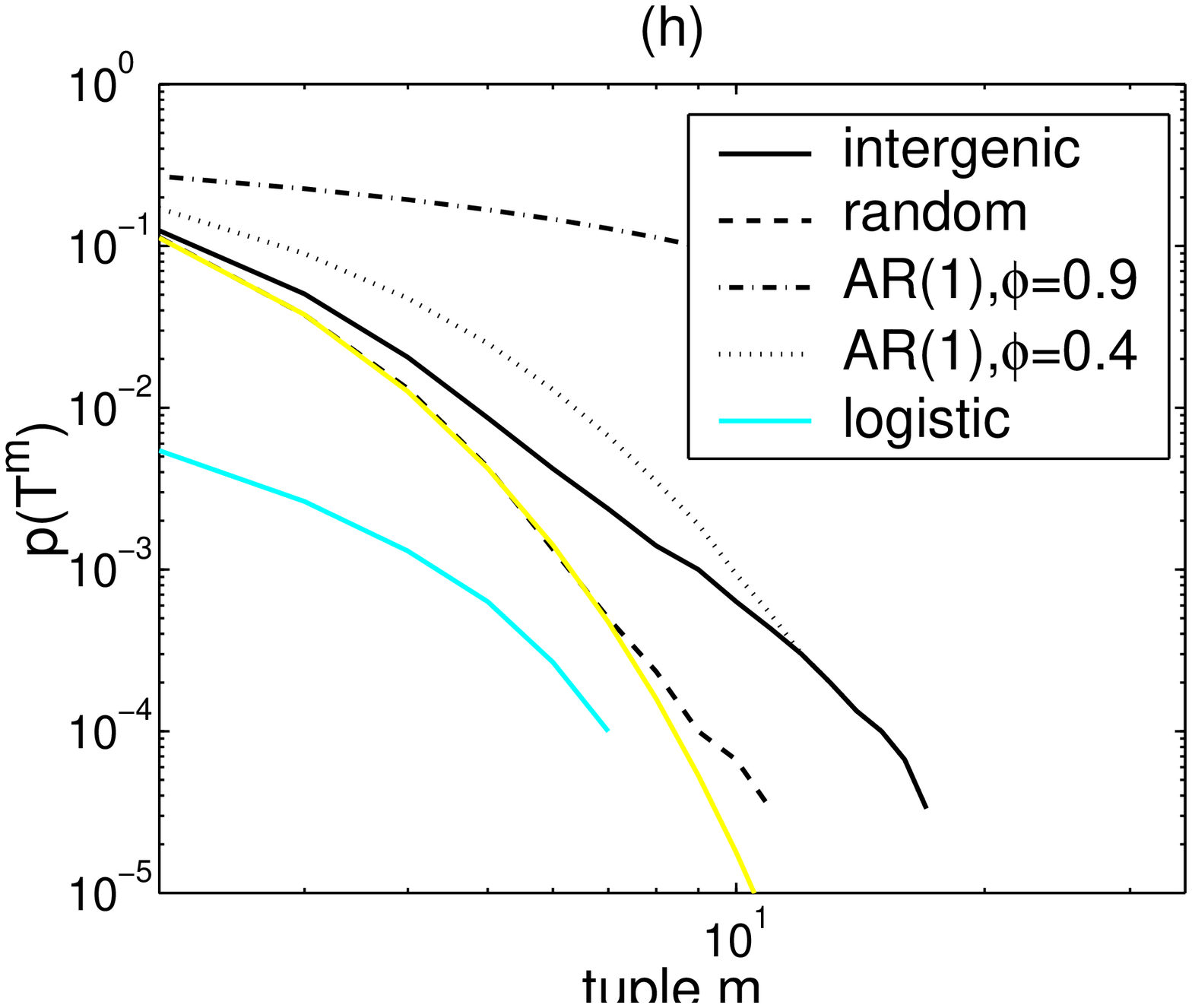}}}
\caption{(a) The graph of the sample probability function of the tuples of 
symbol A evaluated for a gene CAT1 sequence and for four artificial symbolic
sequences with the same base probability as the gene sequence (specified
in the legend). 
All sequences consist of 30000 bases.
The analytic probability function assuming independent sequence is 
also superimposed shown with a light gray curve.
(b) The same as in (a) but for an intergenic CAT1 sequence.
The same results as in (a) and (b) are shown for symbol C in panels 
(c) and (d), for symbol G in
panels (e) and (f), and for symbol T in panels (g) and (h).
}
\label{clustersystems}
\end{figure}

We turn now to compare the two types of DNA sequences to the four 
artificial symbolic sequences. 
The tuple distributions are shown in Fig.~\ref{clustersystems}. 

The tuple distribution for the random sequence is the same as 
the analytic tuple distribution under the assumption of zero correlation
and validates that the use of sequences of 30000 bases gives sufficient
estimation of the probability distribution of the tuples. 

The symbolic sequences from the chaotic logistic map, which is a system of
zero linear correlation but strong nonlinear correlation, have a special
order of the symbols. 
The logistic symbolic sequence based on the base probabilities of the 
gene sequence does not contain a C followed by C or a T followed by T 
(i.e. a point of the numerical time series in a region that is assigned 
to C or T does not map in the same region). 
The same holds for symbols C and G of the logistic symbolic sequence 
based on the base probabilities of the intergenic sequence.
Thus the results from the tuple distribution for symbols C, G and T 
cannot be clearly interpreted. 
For symbol A, the tuple distribution is similar to the weakly 
correlated AR(1) for both types of DNA.  

The strongly correlated AR(1) system has always by far the highest 
probability across all tuples and symbols as expected.
Apparently, the DNA sequences (gene and intergenic) do not contain
strong correlations that would be pronounced as frequent homologous
symbol clusters of any size. 
It seems that even a weakly correlated AR(1) model (for $\phi=0.4$) 
contains more homologous symbol clusters of different size than a gene 
or intergenic sequence.
The estimated probability of tuples tends to be higher for the weakly
correlated AR(1) model than for either of the two DNA sequences.
However, for large tuples the probability decreases slower for the 
intergenic sequence and for very large tuples it is even larger than
the respective probability of the weakly correlated AR(1) symbolic 
sequence, as expected for distributions with long tails. 
This does not hold for the gene sequence and this difference suggests 
that there are stronger long range correlations in the intergenic 
sequence than in the gene sequence.

\subsection{Results from the mutual information}

\begin{figure}[htb] 
\centerline{\hbox{\includegraphics[width=7cm,keepaspectratio]{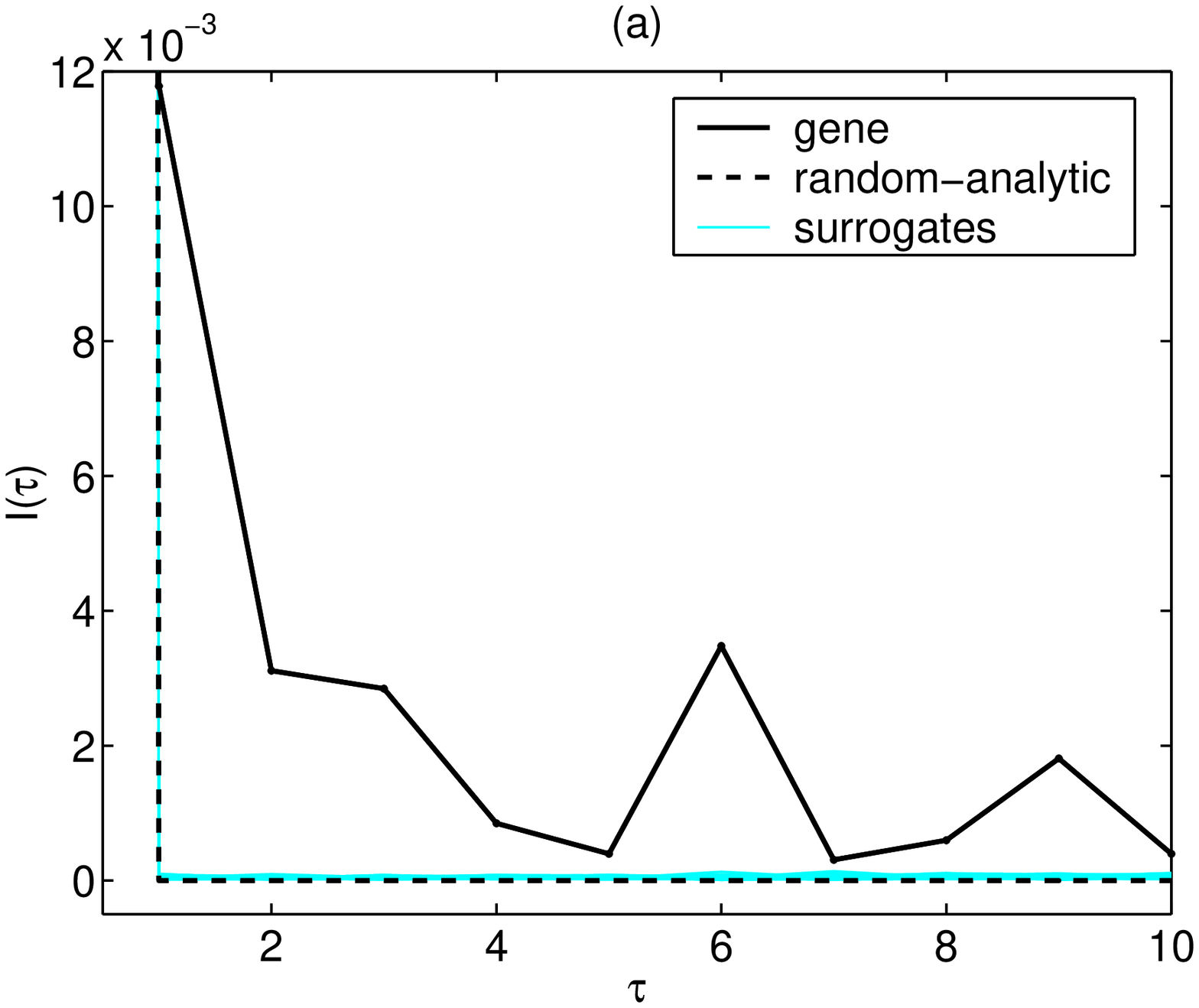}
	\includegraphics[width=7cm,keepaspectratio]{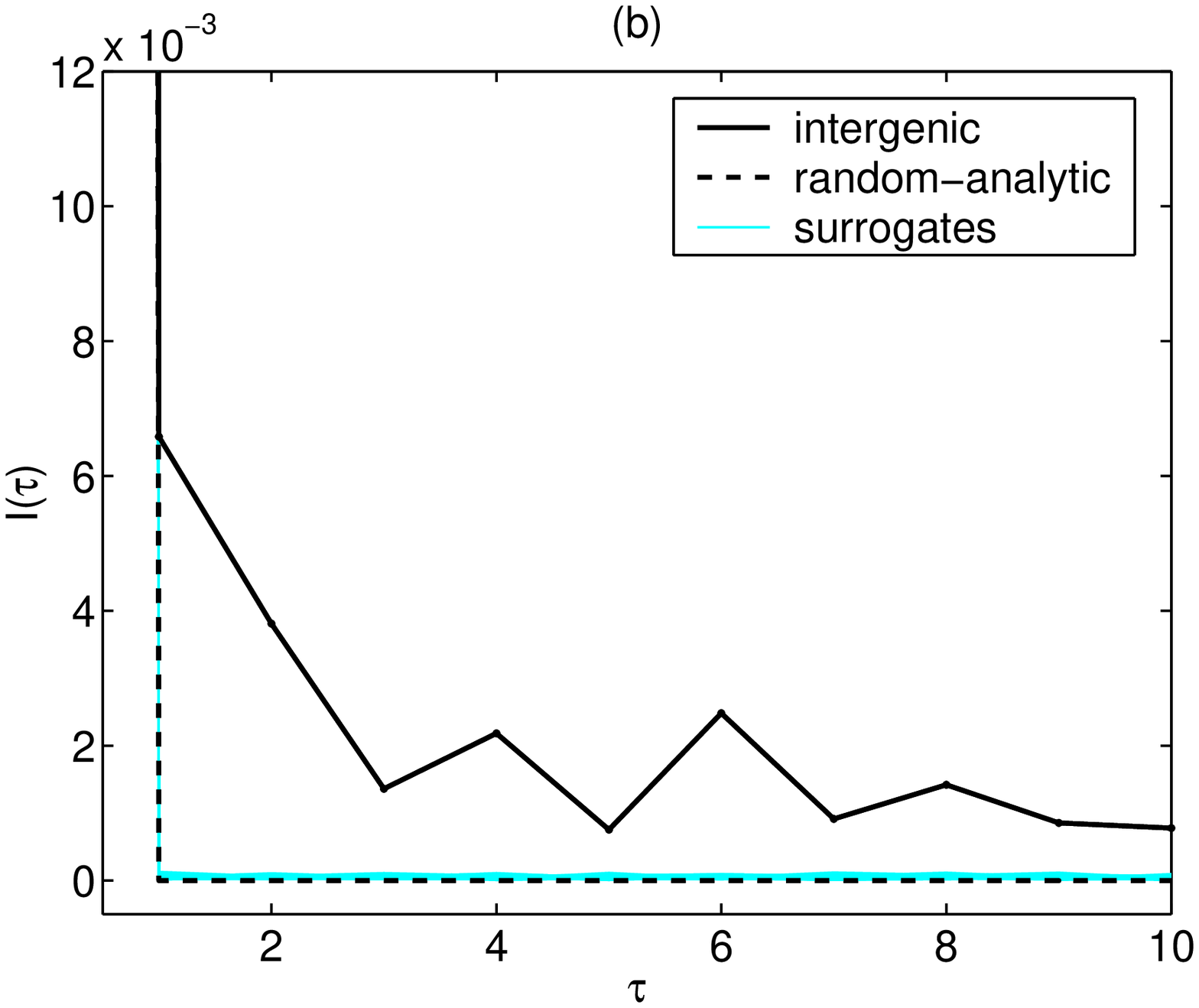}}}
\caption{(a) The graph of the mutual information $I(\tau)$ for displacements 
$\tau=1,\ldots,10$ computed on a gene sequence of 100000 bases of CAT1 
and $M=40$ shuffled surrogates as shown in the legend.
The mutual information for a random sequence is zero for any $\tau$
(this is the curve of ``random-analytic'' in the legend).
(b) The same as in (a) but for an intergenic CAT1 sequence.
}
\label{mutinfosur}
\end{figure}

It has been reported in earlier works that the mutual information function 
has a significantly different functional form in coding and non-coding DNA.
However, there are contradicting reports whether the exons or introns 
have larger mutual information (e.g. see \cite{Grosse00} and 
\cite{Guharay00}).
Here, we study the mutual information of the gene and intergenic CAT1 
sequences together with surrogate symbolic sequences of zero correlations 
and the results are shown in Fig.~\ref{mutinfosur}. 

The simulations with surrogate data simply confirm that the estimation of 
mutual information from a sequence of 100000 is very accurate.
Thus the small values of $I(\tau)$ computed on both types of DNA
sequences are indeed significant and they cannot be attributed to 
deviations from zero due to insufficient statistics. 
So, the immediate conclusion from Fig.~\ref{mutinfosur} is the 
existence of correlations in both the gene and intergenic sequence
(a formal parametric test with surrogate data would give rejection 
of the null hypothesis of zero correlations at very high confidence
levels).  
Further, it does not appear that the intergenic sequence has larger
correlations than the gene sequence. 
To the contrary, $I(1)$ is much larger for gene sequence suggesting
that the general correlations (linear and nonlinear) of adjacent 
symbols is larger in the genes than in the intergenic sequences. 
This is indeed expected because in coding regions symbols up to
the order of 3 code for aminoacids. 
Moreover, from Fig.~\ref{mutinfosur}a it can be seen that there are more
significant correlations in multiples of 3 (gene sequence) while there
are no such features in Fig.~\ref{mutinfosur}b (intergenic sequence).    

\begin{figure}[htb] 
\centerline{\hbox{\includegraphics[width=7cm,keepaspectratio]{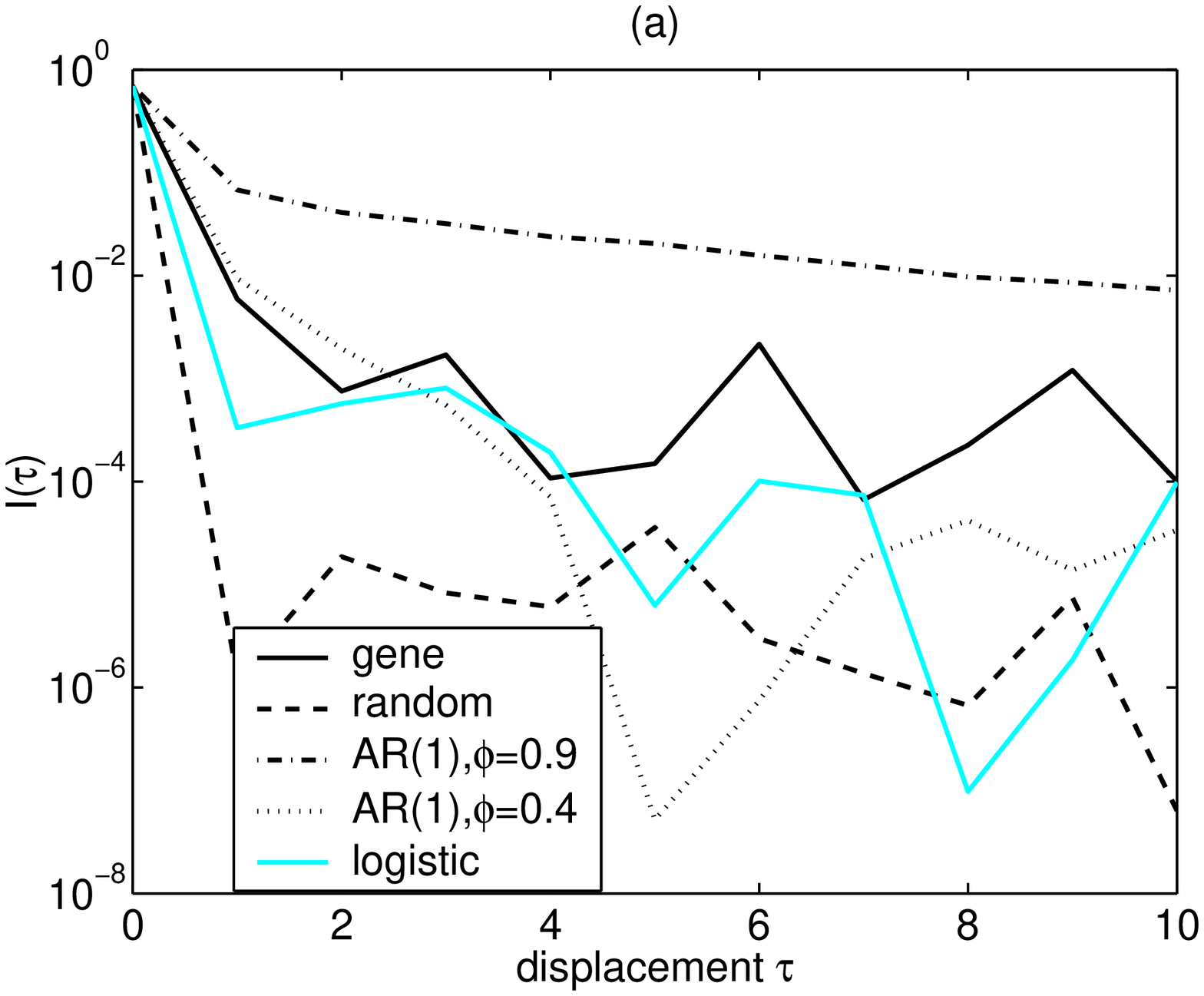}
	\includegraphics[width=7cm,keepaspectratio]{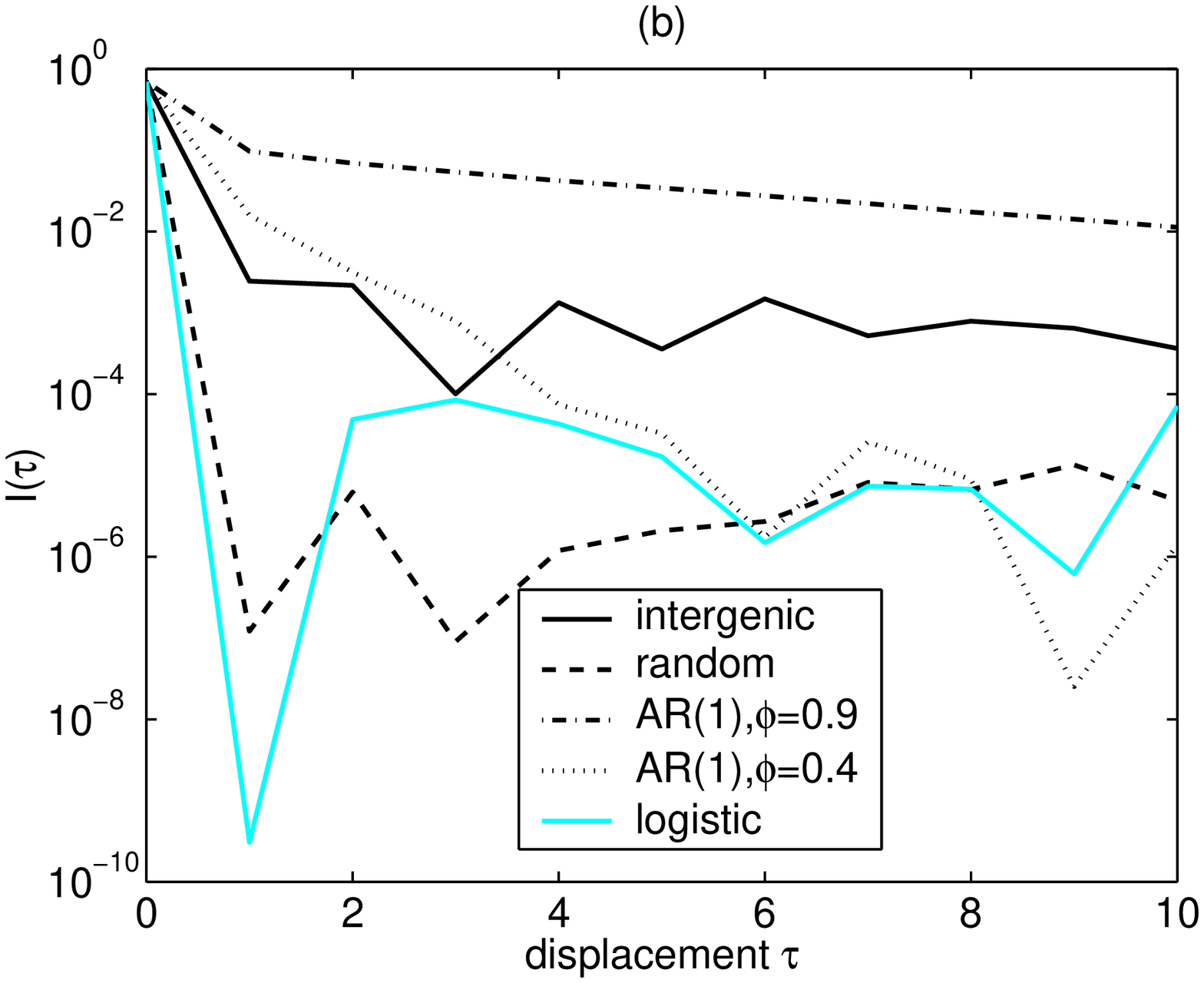}}}
\caption{(a) The graph of the mutual information $I(\tau)$ for displacements 
$\tau=0,1,\ldots,10$ computed on a gene CAT1 sequence of 30000 bases and on
four artificial symbolic sequences as shown in the legend. 
(b) The same as in (a) but for an intergenic CAT1 sequence.
}
\label{mutinfosys}
\end{figure}

Next, we compare $I(\tau)$ from gene and intergenic sequences to 
$I(\tau)$ from the artificial symbolic sequences.
The results are shown in Fig.~\ref{mutinfosys}. 

The random sequence sets the level of the essentially zero $I(\tau)$ 
(the plateau is at about $10^{-6}$).
Similarly to the tuple distribution, the strongly correlated AR(1) 
obtains very high values of $I(\tau)$ and is clearly distinguished from 
all other systems.
The $I(\tau)$ of both gene and intergenic sequences is somehow smaller 
than the $I(\tau)$ for the respective weakly correlated AR(1) sequence and 
larger than the $I(\tau)$ for the logistic symbolic sequence.
So, in terms of mutual information, the DNA sequences would be classified
in-between the weakly correlated AR(1) (for $\phi=0.4$) sequences and the 
logistic symbolic sequences.  

An interesting feature that can be observed from the comparison of 
the two DNA sequences to surrogate data (in Fig.~\ref{mutinfosur}) and 
to other correlated symbolic sequences (in Fig.~\ref{mutinfosys}),
is that the $I(\tau)$ of both DNA sequences does not tend to vanish as 
$\tau$ increases ($I(\tau)>10^{-4}$), which suggests that some correlation 
persists even between symbols that are not spatially close.

\subsection{Results from the identical neighbor fit}

While mutual information involves only two symbols (displaced by $\tau$
spatial units), the identical neighbor fit involves groups of identical
symbol segments of some length $m$ and relates to correlations spanned 
over segments of symbols.
Neighbor fitting and in general nonlinear modeling has not been used 
much in DNA sequence analysis. 
In \cite{Barral00}, a similar technique to the identical neighbor fit was
applied to coding (exons) and non-coding regions and it was found that
coding regions behave as random chains while non-coding regions have 
deterministic structure. 
Our analysis is again different in that the DNA sequences are genes and 
intergenic regions.

\begin{figure}[htb] 
\centerline{\hbox{\includegraphics[width=7cm,keepaspectratio]{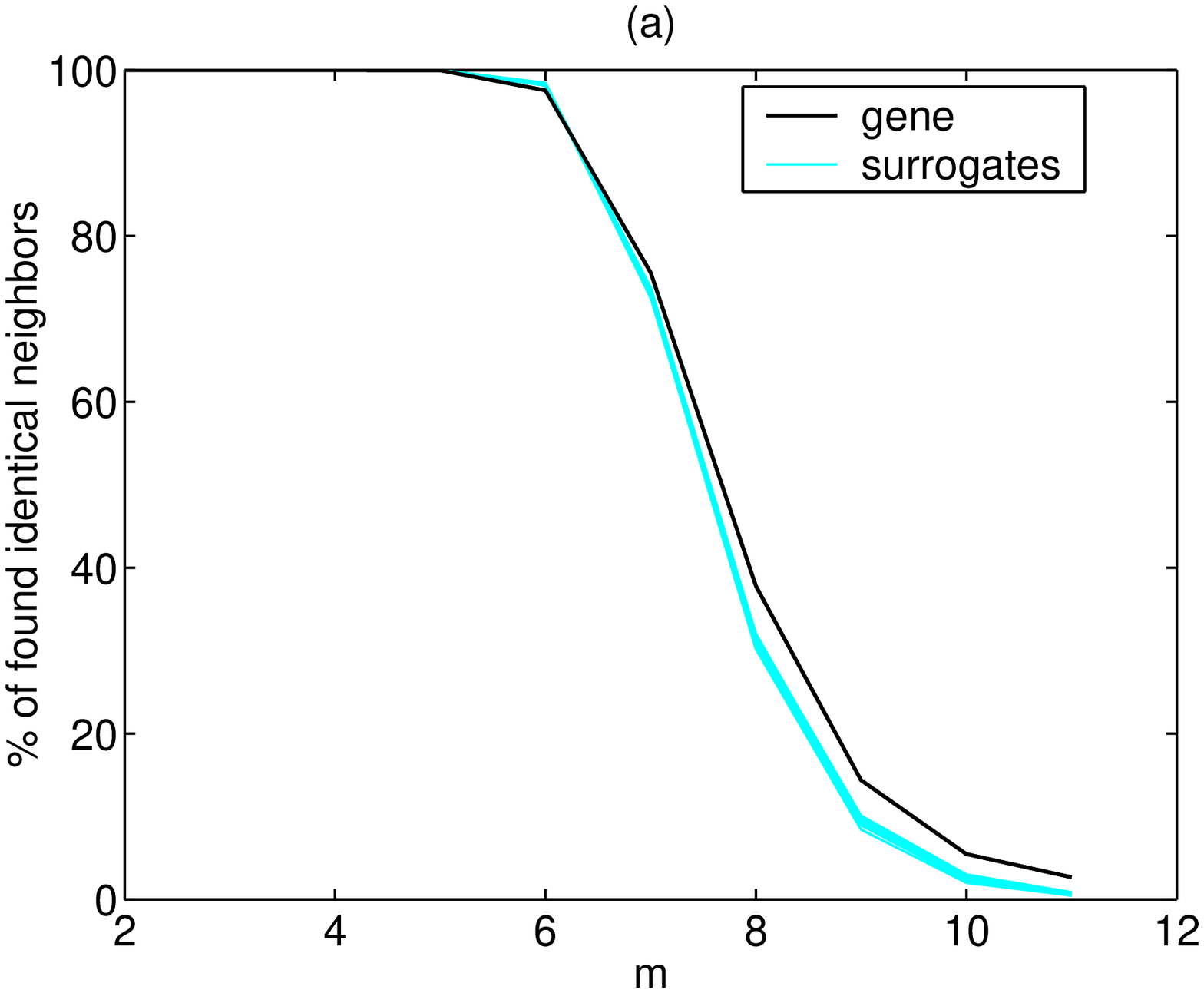}
	\includegraphics[width=7cm,keepaspectratio]{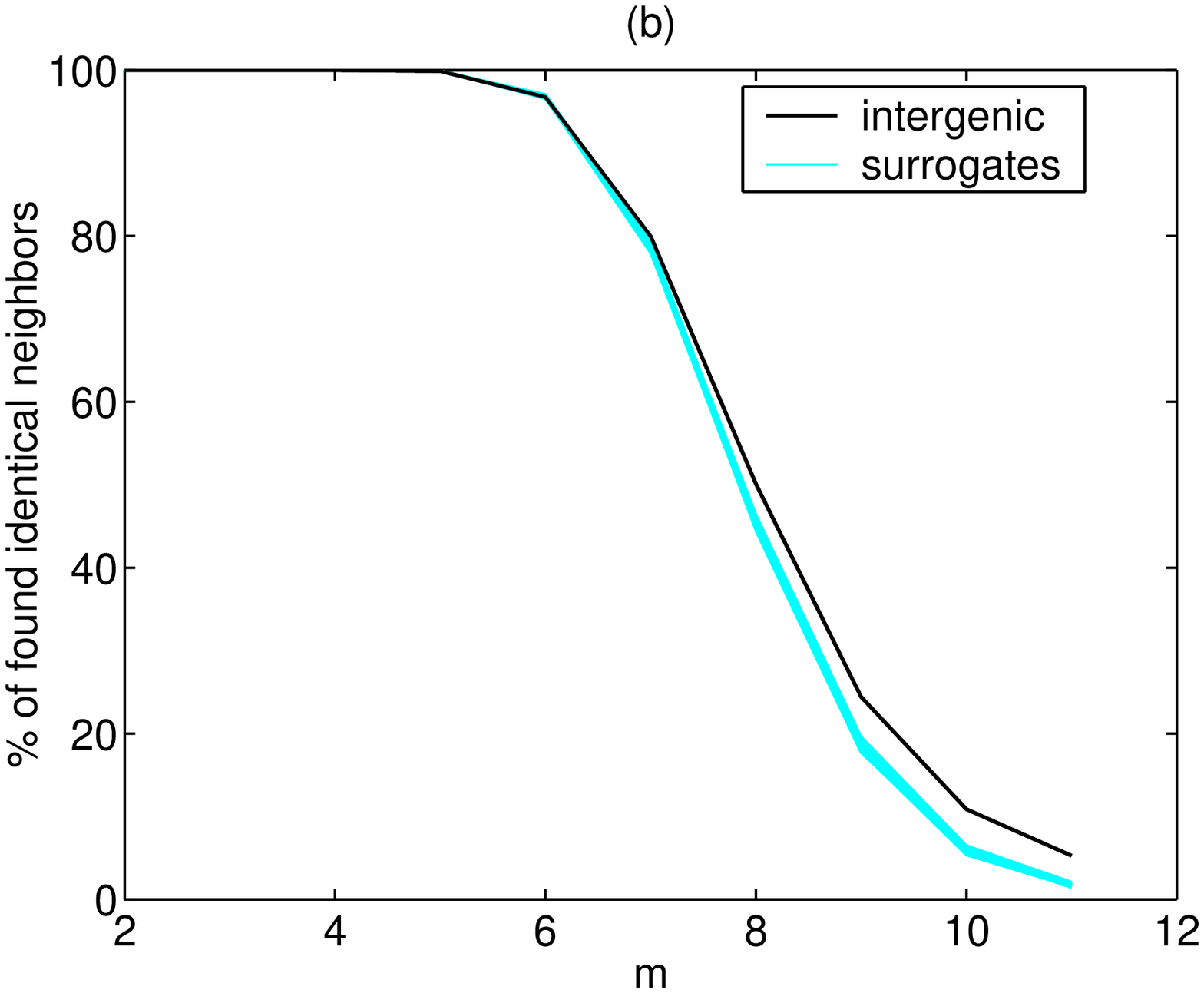}}}
\caption{(a) The percentage of target points for which at least one identical
neighbor was found given as a function of the segment length $m$ and 
computed on a gene CAT1 sequence of 20000 bases and its 40 surrogate 
symbolic sequences.
(b) The same as in (a) but for the intergenic CAT1 sequence.
}
\label{neighsur}
\end{figure}

The $T$ positions ahead identical neighbor fit for a target position $i$ 
relies on the existence of neighbor segments (of some length $m$) in the
sequence that are identical to the current segment. 
Obviously, for very large $m$ there might be no identical neighbors, 
depending also on the length of the sequence.
So, the existence of identical neighbors for large $m$ may signify a 
special deterministic structure or correlation in the sequence.
In Fig.~\ref{neighsur} we show for the DNA sequences and their
surrogates the proportion of target positions over all possible positions 
in the sequence for which at least one identical neighbor was found 
(and prediction could be made). 
We observe that for large segments, say $m>8$, the percentage of existing
identical neighbors decreases to zero, but the decrease is slower for the
DNA sequences (gene and intergenic regions) than for their counterpart
surrogate sequences.
This indicates that there is some form of organization of the symbols in
the DNA sequence, so that particular combinations of $m$ symbols occur
more often than if the chain of symbols were completely random.

\begin{figure}[htb] 
\centerline{\hbox{\includegraphics[width=7cm,keepaspectratio]{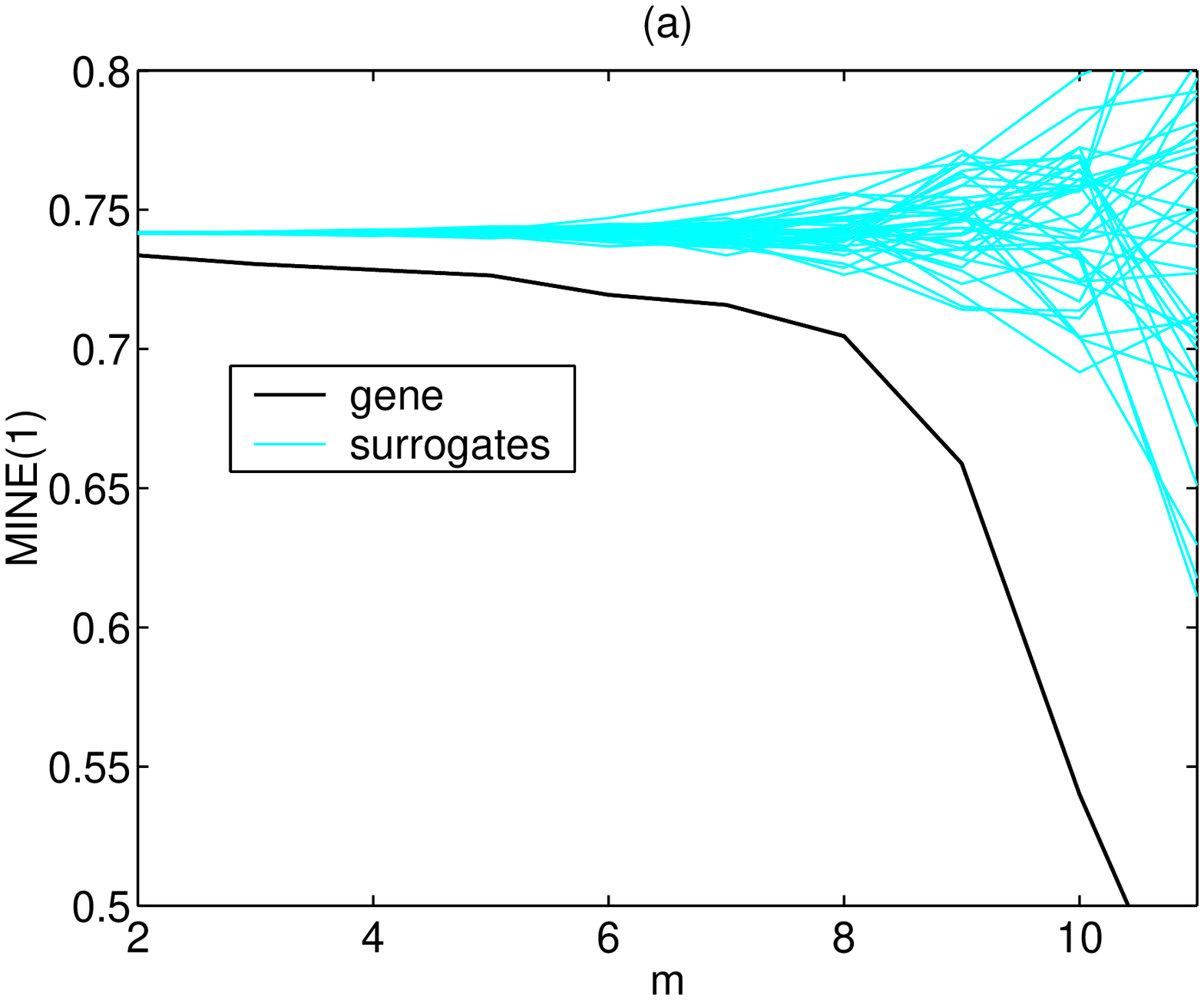}
	\includegraphics[width=7cm,keepaspectratio]{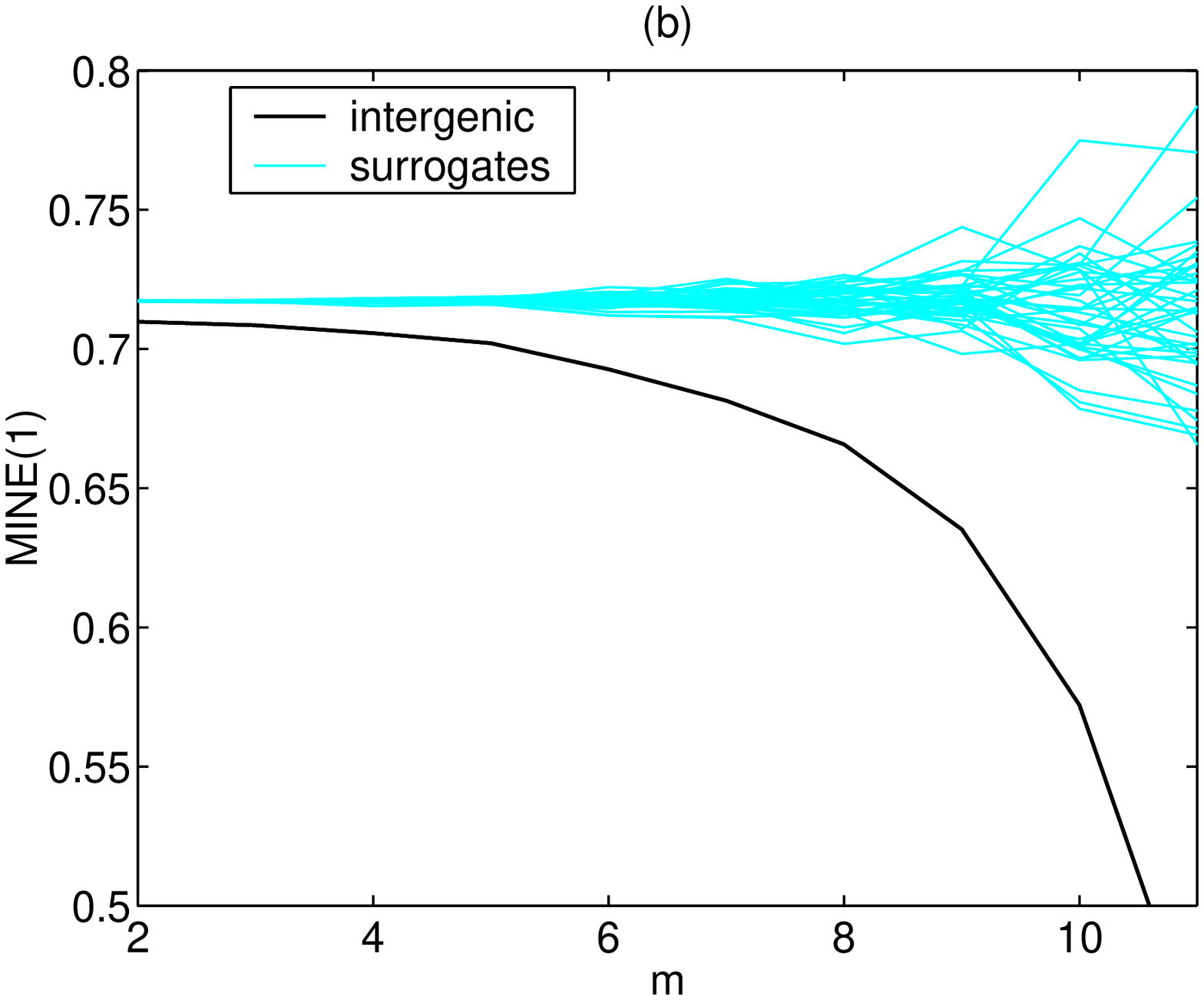}}}
\centerline{\hbox{\includegraphics[width=7cm,keepaspectratio]{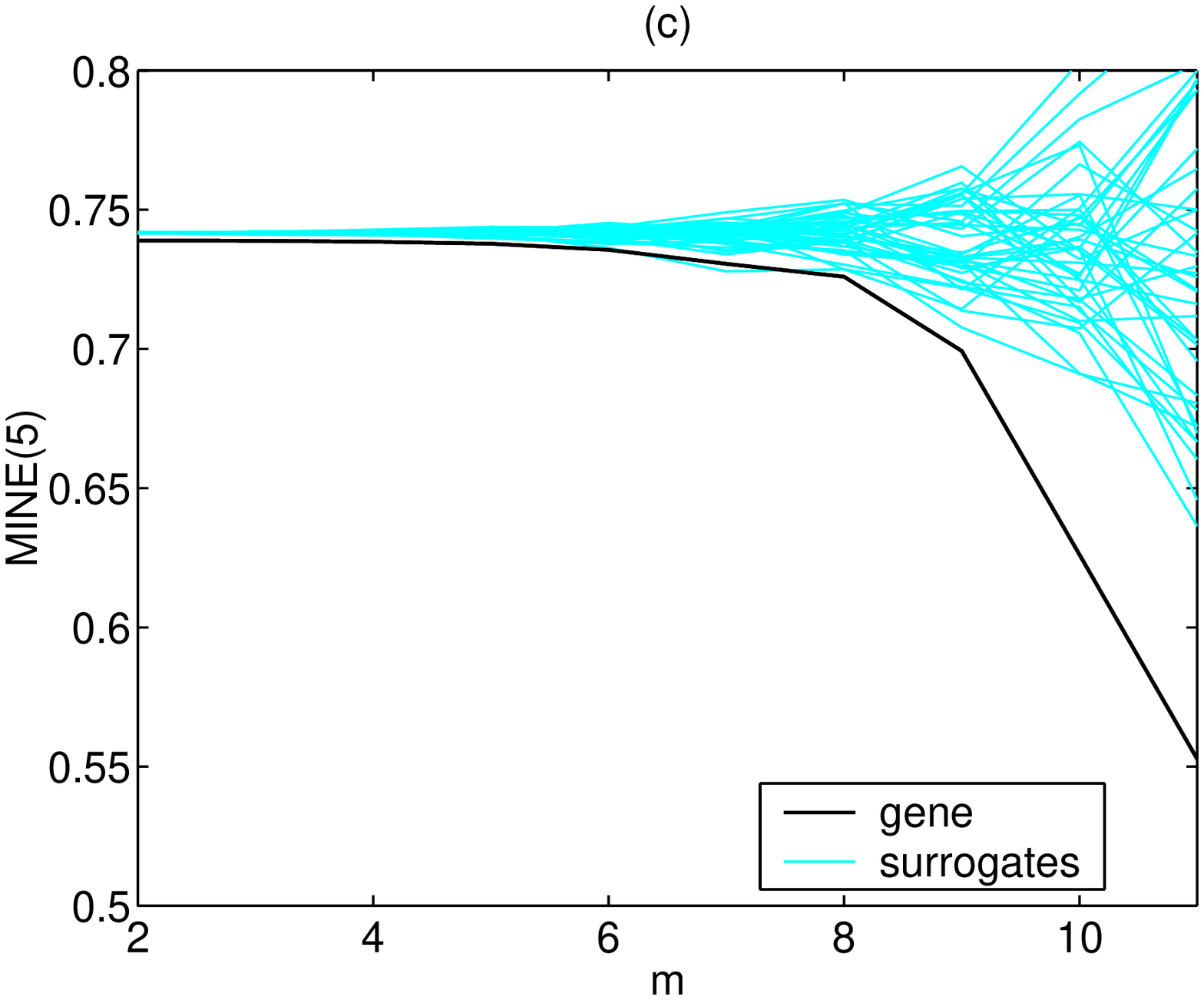}
	\includegraphics[width=7cm,keepaspectratio]{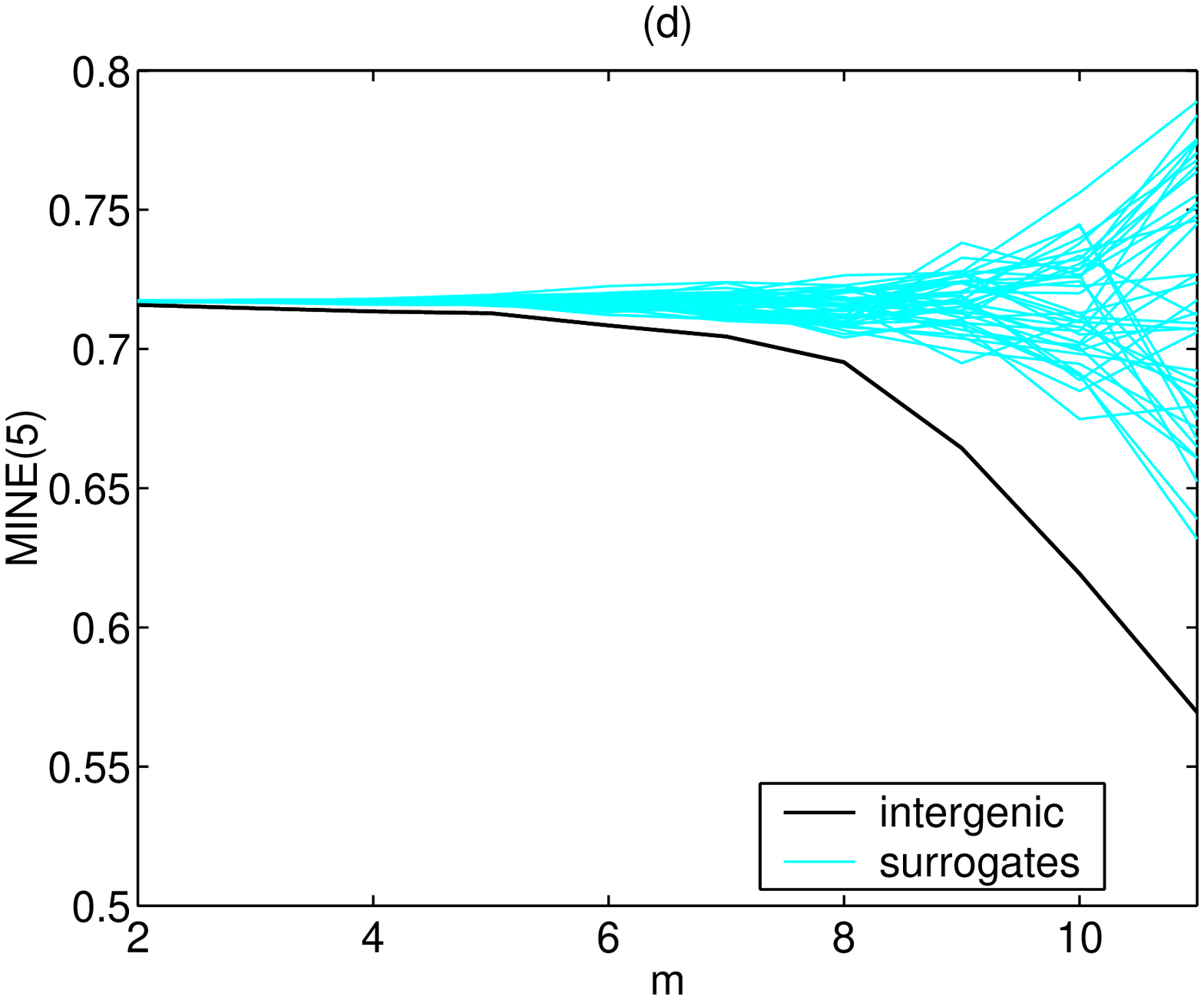}}}
\caption{(a) The one step ahead fit error MIME (T=1) as a function of 
the segment length $m$ computed on a gene CAT1 sequence of 20000 bases 
and its 40 surrogate symbolic sequences.
(b) The same as (a) but for the intergenic CAT1 sequence.
(c) The same as (a) but for 5 step ahead fit (T=5).
(d) The same as (b) but for 5 step ahead fit.
}
\label{MIMEsur}
\end{figure}

Next, we investigate whether the prediction based on identical
neighbors is better for the DNA sequences than for their surrogates.
As shown in Fig.~\ref{MIMEsur}, the differences are indeed significant, 
they increase with $m$ and persist even when predicting multi--steps 
ahead.       
In particular, for large $m$ the fit error measure MIME falls dramatically
for the DNA sequences, whereas for the surrogates, MIME is the same on average 
for all $m$ (as expected) but has larger variance due to the decrease of
identical neighbor statistics with $m$.
Note that the level of MIME of the surrogates is different for the gene
and intergenic sequence because it is determined by the base probabilities
which are different for the two DNA sequences (see (\ref{eq:maxmime})).
From Fig.~\ref{MIMEsur}, we cannot draw confidently a different signature 
for the two DNA sequences in terms of identical neighbor prediction, though
it seems that the MIME of the intergenic sequence deviates more from the 
MIME of the respective surrogates for larger $m$ than it does for the 
gene sequence.
The results with the weighted MINE (WMINE) were similar.

It should be stressed that the significant differences shown in 
Fig.~\ref{MIMEsur} 
for $T=1$ and $T=5$ were also observed for a range of prediction steps $T$ 
up to 50 (not shown here) when $m$ is small (at the level of 4).
This indicates that given the occurrence of a particular small 
sequence of 3 to 5 nucleotides the probability of observing a specific
nucleotide many positions ahead in the sequence is non-trivial for 
both the genes and intergenic sequences. 
However, this probability does not seem to differ systematically between
genes and intergenic sequences.
 
\begin{figure}[htb] 
\centerline{\hbox{\includegraphics[width=7cm,keepaspectratio]{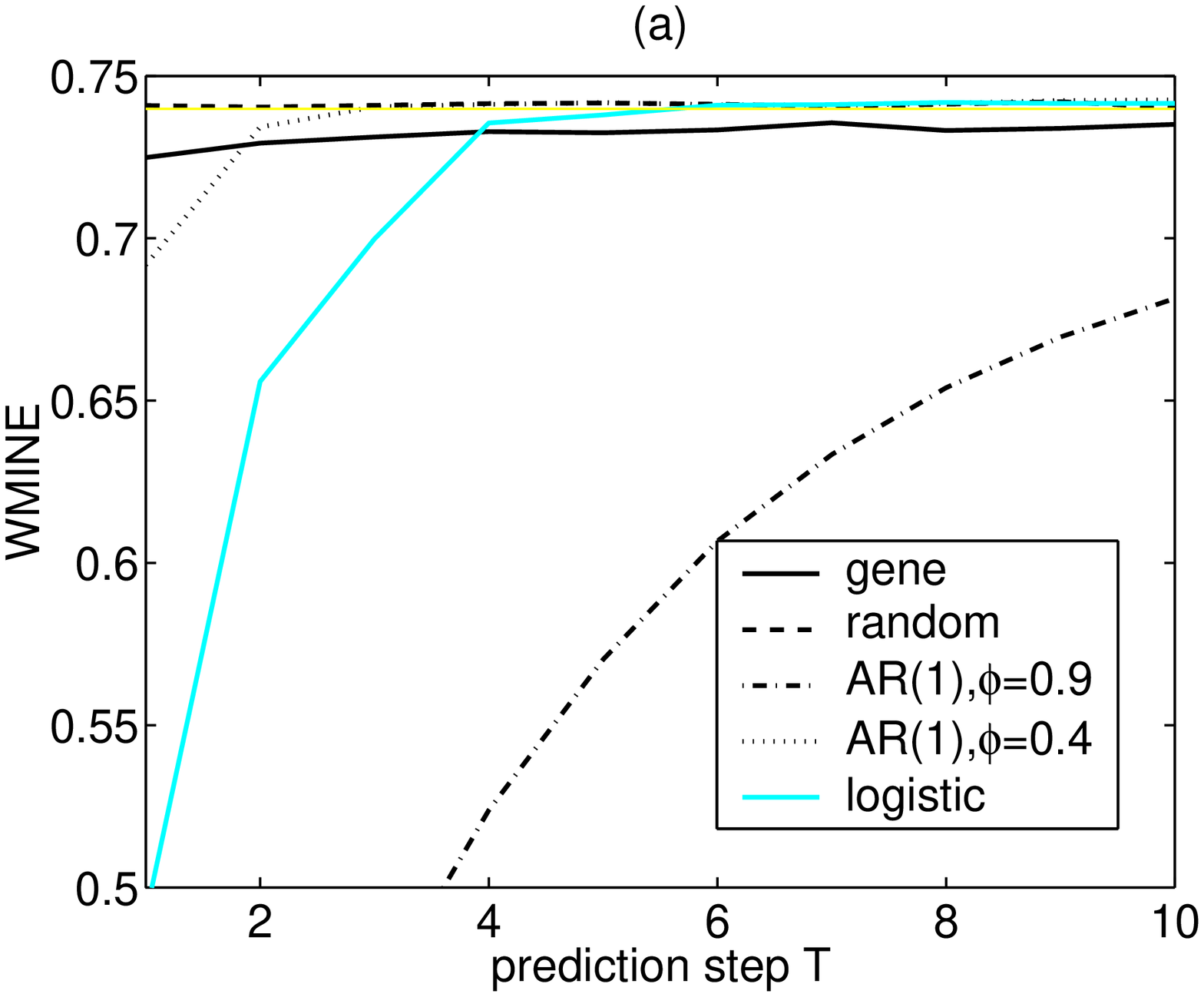}
	\includegraphics[width=7cm,keepaspectratio]{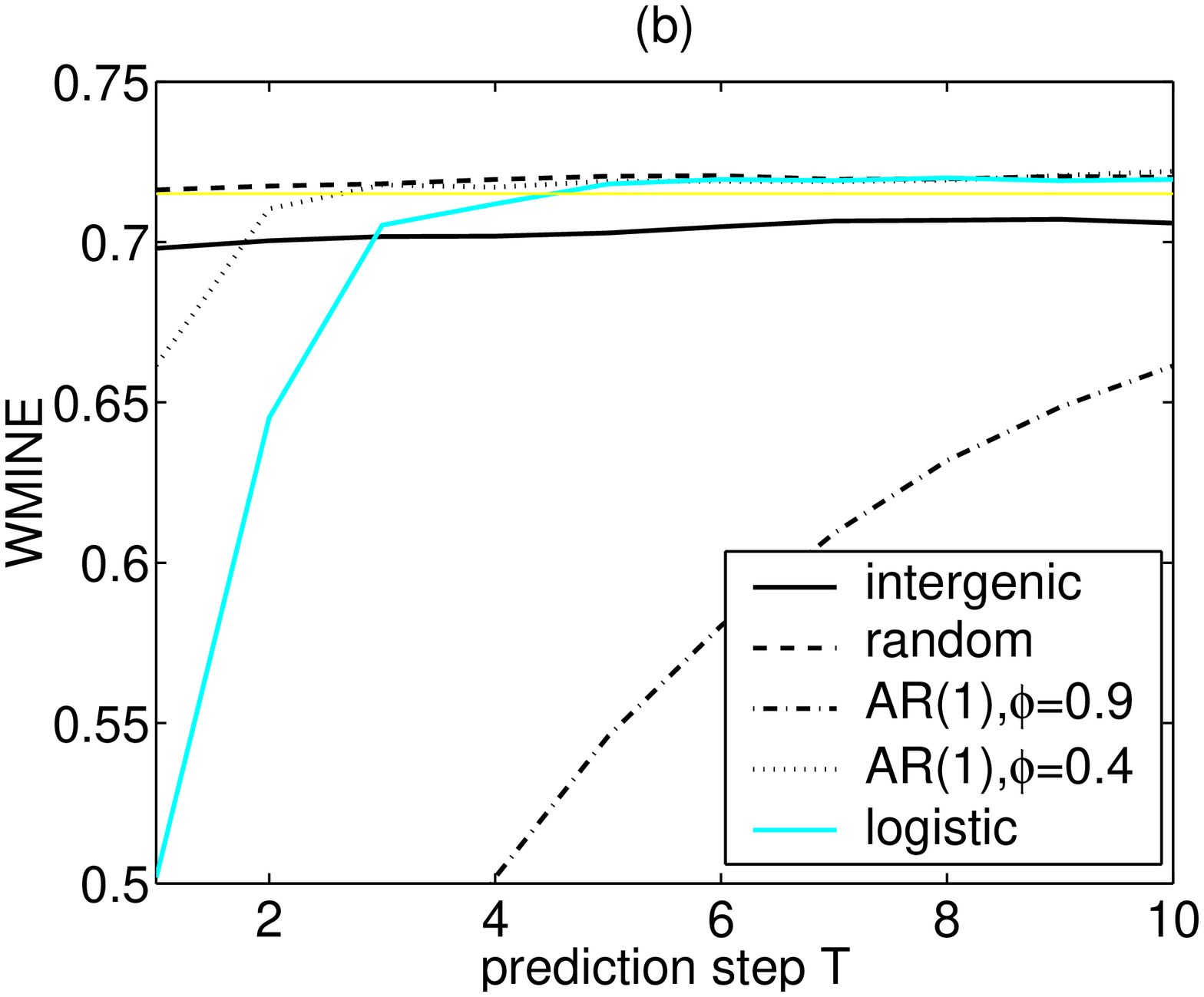}}}
\caption{(a) The fit error WMIME($T$) as a function of the prediction step
$T$ for segment length $m=5$ computed on a gene CAT1 sequence of 30000 
bases and the respective four artificial symbolic sequences, as shown in 
the legend.
(b) The same as (a) but for the intergenic CAT1 sequence.
}
\label{MIMEsys}
\end{figure}

Finally, we compare using the identical neighbor fit the DNA sequences
to the artificial symbolic sequences. 
We show in Fig.~\ref{MIMEsys} the results of WMINE(T) for $T=1,\ldots,10$ 
and $m=5$ on all the sequences. 
The maximum WMIME (and MIME) for a symbolic sequence (see (\ref{eq:maxmime}))
is essentially the same as the WMIME for the random symbolic sequence.
The one step ahead prediction ($T=1$) for the DNA sequences is worse than 
all but the random sequence. 
When the prediction is made for many steps ahead, say $T>4$, the WMIME for 
the logistic sequence and the weakly correlated AR(1) converges to the 
maximum WMIME (that correspond to the MIME of a random sequence) while 
the WMIME for the DNA sequences does not change much and remains at a lower
than the maximum WMIME level.
It should be noted that the WMIME for the strongly correlated AR(1) is 
much smaller.

\section{Discussion}
\label{Discussion} 

In the current study we use three statistical methods to establish
differences and similarities between genes and intergenic regions
in Chromosome 1 of {\em Arabidopsis Thaliana} as well as artificially
generated symbolic sequences from well-known systems.

Our analysis could discriminate the correlation structure of the 
different symbolic sequences.
The following results were obtained with the three measures of tuple 
probability distribution, mutual information and identical neighbor fit:
a) Both genes and intergenic DNA sequences have significant correlations
as compared to the respective random surrogate data (sequences having
no correlation structure). 
b) The intergenic regions tend to be more correlated than genes
for larger scales of displacement (addressed by $\tau$ for mutual
information and by $m$ for tuple probability distribution and identical
neighbor fit).
c) Compared to simulated systems, the correlations of both gene and 
intergenic DNA sequences are close to those of an AR(1) model with 
autocorrelation function $r(\tau)=0.4^{\tau}$, (note that the numerical 
time series of the system is suitably discretized to four symbols, so 
that the probability distribution of the symbols is identical to this 
of the respective DNA sequence).
d) The resemblance of DNA correlation to the correlation of AR(1) 
holds only for small scales of displacement (according to the measure,
$\tau$ or $m$ smaller than 4), while for larger displacements the
DNA retains significant correlation. 

Certainly, the aim of the parallel studies of  the DNA sequence of 
{\em Arabidopsis Thaliana} and of the different simulated systems was
not to find the best suited system for this natural sequence in terms of
correlation structure, but rather to compare the correlation signature 
of DNA sequence to this of other symbolic sequences derived from well known
systems. 
In future efforts, we intend to investigate further whether known 
deterministic or stochastic systems can give rise to symbolic sequences 
that imitate the correlation structure of the DNA sequences.

The DNA sequence of genes and intergenic regions used in this analysis
is a concatenation of individual genes and intergenic segments, 
both not exceeding a couple of thousand bases on average.
Thus long-range correlations cannot be properly estimated, moderate-range
correlations (at the level of hundreds) are underestimated and in 
general the estimated correlations are reduced due to the frequent 
discontinuity in both series. 
Therefore, we found useful to use the surrogate data as a reference of 
complete lack of correlations and assess the departure from this level.
Moreover, in our analysis we concentrated basically on medium scale 
correlations (e.g. windows of about 10 bases). 
  
In other studies concerning DNA statistical analysis
\cite{Dokholyan97,Stanley99,Almirantis99,Provata02}, long range
correlations are found in non-coding DNA while coding DNA presents
more short range correlated features.
In our work the existence of correlations is partly masked since: a) gene regions are mixed
containing coding and non-coding parts, and b) the DNA sequences used 
in the analysis consist of pieces of limited sizes put together.

Regarding the particular DNA sequence used in the analysis, it should be
stressed that although {\em arabidopsis thaliana} belongs to the general 
category of higher eucaryotes, it belongs to the sub-category of dicot
and contains high percentage of coding DNA (approximately 50\%).
In this respect, it resembles more to the lower organisms in which the 
coding is prevailing.
This is why we have found in several cases strong levels of resemblance between 
genes and intergenic regions. 
For example, in Fig.\,\ref{clustersur}j there seem to be strong large 
size T-clusters within the gene regions which can be attributed to the 
introns of the corresponding gene.

Other plant sequences, such as the rice genome (now nearly completely
sequenced), which is a monocot and has characteristics closer to higher
organisms, need to be investigated and compared with the results on
{\em arabidopsis thaliana}.
Also, the same analysis need to be carried out for different classes
of organisms (animals, lower eycaryotes, etc.), so that similarities
and differences between various categories of organisms could be
assessed.

While all methods discriminate significantly the DNA sequences (genes
and intergenic regions) from the respective random surrogate data 
it remains still to be investigated whether this discrimination holds
for small data sizes at the level of 1000 bases.
Some preliminary results showed that the identical neighbor fit has
better power at discriminating DNA segments of 1000 bases from surrogate
data for small $m$ and $T$ but more systematic analysis is required on
this. 




\clearpage

\begin{thebibliography}{10}

\bibitem{Peng92}
C.-K. Peng, S.~V. Buldyrev, A.~L. Goldberger, S.~Havlin, F.~Sciortino,
  M.~Simons, and H.~E. Stanley.
\newblock Long-range correlation in nucleotide-sequences.
\newblock {\em Nature}, 365(6365):168 -- 170, 1992.

\bibitem{Buldyrev95}
S.~V. Buldyrev, A.~L. Goldberger, S.~Havlin, R.~N. Mantegna, M.~E. Matsa, C.-K.
  Peng, M.~Simons, and H.~E. Stanley.
\newblock Long-range correlation-properties of coding and noncoding
  {DNA}-sequences -- {G}enbank analysis.
\newblock {\em Physical Review E}, 51(5):5084 -- 5091, 1995.

\bibitem{Peng94}
C.-K. Peng, S.~V. Buldyrev, S.~Havlin, M.~Simons, H.~E. Stanley, and A.~L.
  Goldberger.
\newblock Mosaic organization of {DNA} nucleotides.
\newblock {\em Physical Review E}, 49(2):1685 -- 1689, 1994.

\bibitem{Ebeling91}
W~Ebeling and G.~Nicolis.
\newblock Entropy of symbolic sequences: the role of correlations.
\newblock {\em Europhysics Letters}, 14(3):191 -- 196, 1991.

\bibitem{Stanley99b}
H.~E. Stanley, S.~V. Buldyrev, A.~L. Goldberger, S.~Havlin, C.-K. Peng, and
  M.~Simons.
\newblock Scaling features of noncoding {DNA}.
\newblock {\em Physica A}, 273:1 -- 18, 1999.

\bibitem{Herzel97}
H.~Herzel and I.~Grosse.
\newblock Correlations in {DNA} sequences: The role of protein coding segments.
\newblock {\em Physical Review E}, 55(1):800 -- 810, 1997.

\bibitem{Grosse00}
I.~Grosse, H.~Herzel, S.~V. Buldyrev, and H.~E. Stanley.
\newblock Species independence of mutual information in coding and noncoding
  {DNA}.
\newblock {\em Physical Review E}, 61(5):5624 -- 5628, 2000.

\bibitem{Guharay00}
B.~R. Guharay, S.~Hunt, J.~A. Yorke, and O.~R. White.
\newblock Correlations in {DNA} sequences across the three domains of life.
\newblock {\em Physica D}, 146:388 -- 396, 2000.

\bibitem{Buldyrev98}
S.~V. Buldyrev, N.~V. Dokholyan, A.~L. Goldberger, S.~Havlin, C.-K. Peng, H.~E.
  Stanley, and G.~M. Viswanathan.
\newblock Analysis of {DNA} sequences using methods of statistical physics.
\newblock {\em Physica A}, 249:430 -- 438, 1998.

\bibitem{Dokholyan97}
N.~V. Dokholyan, S.~V. Buldyrev, S.~Havlin, and H.~E. Stanley.
\newblock Distribution of base pair repeats in coding and noncoding {DNA}
  sequences.
\newblock {\em Physical Review Letters}, 79(25):5182 -- 5185, 1997.

\bibitem{Stanley99}
R.~H.~R. Stanley, N.~V. Dokholyan, S.~V. Buldyrev, S.~Havlin, and H.~E.
  Stanley.
\newblock Clustering of identical oligomers in coding and noncoding {DNA}
  sequences.
\newblock {\em Journal of Biomolecular Structure and Dynamics}, 17(1):79 -- 87,
  1999.

\bibitem{Almirantis99}
Y.~Almirantis and A.~Provata.
\newblock Long and short range correlations in genome organization.
\newblock {\em Journal of Statistical Physics}, 97:233 -- 262, 1999.

\bibitem{Provata02}
A.~Provata and Y.~Almirantis.
\newblock Statistical dynamics of clustering in the genome structure.
\newblock {\em Journal of Statistical Physics}, 106(1/2):23 -- 55, 2002.

\bibitem{Ebeling87}
W.~Ebeling, R.~Feistel, and H.~Herzel.
\newblock Dynamics and complexity of biomolecules.
\newblock {\em Physica Scripta}, 35(5):761 -- 768, 1987.

\bibitem{Schreiber99b}
T.~Schreiber and A.~Schmitz.
\newblock Surrogate time series.
\newblock {\em Physica D}, 142(3-4):346 -- 382, 2000.

\bibitem{Kugiumtzis01a}
D.~Kugiumtzis.
\newblock Surrogate data test on time series.
\newblock In A.~Soofi and L.~Cao, editors, {\em Modelling and Forecasting
  Financial Data, Techniques of Nonlinear Dynamics}, chapter~12, pages 267 --
  282. Kluwer Academic Publishers, 2002.

\bibitem{Shannon49}
C.~E. Shannon and W.~Weaver.
\newblock {\em The Mathematical Theory of Communication}.
\newblock University of Illinois Press, Urbana, 1949.

\bibitem{Fraser86}
A.~M. Fraser and H.~Swinney.
\newblock Independent coordinates for strange attractors from mutual
  information.
\newblock {\em Physical Review A}, 33:1134 -- 1140, 1986.

\bibitem{Kantz97}
H.~Kantz and T.~Schreiber.
\newblock {\em Nonlinear Time Series Analysis}.
\newblock Cambridge University Press, Cambridge, 1997.

\bibitem{Barral00}
J.~P. Barral, A.~Hasmy, J.~Jim\'{e}nez, and A.~Marcano.
\newblock Nonlinear modeling technique for the analysis of {DNA} chains.
\newblock {\em Physical Review E}, 61(2):1812 -- 1815, 2000.

\end{thebibliography}
\end{document}